\documentclass[10pt,a4paper,final]{iopart}

\usepackage{iopams}
\usepackage{stackrel}
\usepackage{fullpage}
\usepackage{graphicx}
\usepackage[breaklinks=true,colorlinks=true,linkcolor=blue,urlcolor=blue,citecolor=blue]{hyperref}
\usepackage[capitalize,nameinlink,noabbrev]{cleveref}
\usepackage[utf8]{inputenc}
\usepackage[english]{babel}
\usepackage{enumitem}
\usepackage[toc,page]{appendix}
\usepackage{bm}
\usepackage{array}
\usepackage{mdframed}
\usepackage{braket}
\renewcommand\bra[1]{{\langle{#1}|}}
\renewcommand\ket[1]{{|{#1}\rangle}}

\usepackage{comment}
\usepackage{multirow}
\usepackage[format=hang,font=small,labelfont=bf,textfont=sf,figurename=Fig.]{caption}[2013/02/03]
\usepackage[margin=0.2cm]{subcaption}
\usepackage[usenames,dvipsnames,svgnames,table]{xcolor}
\usepackage{breqn}
\usepackage{cite}

\newtheorem{thm}{Theorem}
\newtheorem{lemma}{Lemma}

\eqnobysec

\begin{document}
	\renewcommand{\equationautorefname}{Eq.} 
	\renewcommand{\figureautorefname}{figure} 
	\renewcommand{\chapterautorefname}{Ch.} 
	\renewcommand{\sectionautorefname}{Sec.} 
	\renewcommand{\subsectionautorefname}{Subsec.} 
	\newcommand{\Mod}[1]{\ (\mathrm{mod}\ #1)}

\title{Conference key agreement with single-photon interference}

\author{Federico Grasselli, Hermann Kampermann and Dagmar Bru\ss}
\address{Institut für Theoretische Physik III, Heinrich-Heine-Universität Düsseldorf, Universitätsstraße 1, D-40225 Düsseldorf, Germany}
\ead{federico.grasselli@hhu.de} 
\begin{abstract}
The intense research activity on Twin-Field (TF) quantum key distribution (QKD) is motivated by the fact that two users can establish a secret key by relying on single-photon interference in an untrusted node. Thanks to this feature, variants of the protocol have been proven to beat the point-to-point private capacity of a lossy quantum channel. Here we generalize the main idea of the TF-QKD protocol introduced by Curty \textit{et al.} to the multipartite scenario, by devising a conference key agreement (CKA) where the users simultaneously distill a secret conference key through single-photon interference. The new CKA is better suited to high-loss scenarios than previous multipartite QKD schemes and it employs for the first time a $W$-class state as its entanglement resource. We prove the protocol's security in the finite-key regime and under general attacks. We also compare its performance with the iterative use of bipartite QKD protocols and show that our truly multipartite scheme can be advantageous, depending on the loss and on the state preparation.
\end{abstract}

The most mature and developed application of quantum communication \cite{quantum-communication-review,Kimble} is certainly quantum key distribution (QKD) \cite{BB84,E91,Scarani-review,Curty-review,Diamanti-review,Pirandola-review}. The majority of the QKD protocols proposed so far involve just two end-users, Alice and Bob, who want to establish a secret shared key. Nowadays there is a vibrant research towards protocols which are proven to be secure in the most adversarial situation possible (i.e. reducing the assumption on the devices) \cite{Curty-MDIQKD,Abruzzo-MDIQKD, mem-assisited-MDIQKD,Azuma-intercityQKD,DIQKD1,DIQKD2}, but at the same time are also implementable with today's technology \cite{404km,421km,free-space-QKD1,free-space-QKD2}. In this context, a protocol which recently received great attention is the Twin-Field (TF) QKD protocol originally proposed by Lucamarini \textit{et al.} \cite{Lucamarini-TF}, further developed to prove its security \cite{Tamaki-security-TF, Ma-security-TF, Cui-security-TF, Lutkenhaus-security-TF, Curty-security-TF,Wang-security-TF,Grasselli-Curty-TF,Grasselli-Navarrete-TF} and experimentally implemented \cite{experiment-chinese,experiment-Toshiba,experiment-Toronto,experiment-Wang}. Indeed, the TF-QKD protocol relies only on single-photon interference occurring in an untrusted node, making it a measurement-device-independent (MDI) QKD protocol capable of overcoming the repeaterless bounds \cite{Takeoka,PLOB}.\\
In a scenario where several users are required to share a common secret key, one can for instance perform bipartite QKD protocols between pairs of users and then use the secret keys established in this way to encode the final common secret key. Alternatively, one can perform a truly multipartite QKD scheme --also known as conference key agreement (CKA)-- whose purpose is to deliver the \textit{same} secret key to \textit{all} the parties involved in the protocol \cite{Epping,Ribeiro,Grasselli,Jo,Pirandola-CV-CKA}. In order to accomplish such a task, a resource which seems necessary is the multipartite Greenberger-Horne-Zeilinger (GHZ) state \cite{Epping,Ribeiro,Grasselli,Jo} or a multipartite private state --a ``twisted'' version of the GHZ state \cite{Horodecki,Azuma-broadcast-network}.\\
In this work we introduce a CKA which exploits for the first time the multipartite entanglement of a $W$-class state \cite{Wstate}, in order to deliver the same secret key to all users. Despite having a number of users involved, the scheme relies on single-photon interference in an untrusted node and it is inspired by the bipartite TF-QKD protocol by Curty \textit{et al.} \cite{Curty-security-TF}. We prove the security of our CKA in the finite-key scenario, allowing Eve to perform the most general attacks (coherent attacks) on the transmitted signals. We compare the performance of our genuinely multipartite QKD scheme with the iterative use of bipartite QKD protocols, both in the asymptotic regime and in the finite-key regime. In doing so, we show that performing a truly multipartite scheme can yield a higher secret key rate, depending on the loss and on the state preparation.\\
The paper is structured as follows. In \autoref{CKA-protocol} we present the CKA based on single-photon interference, while in \autoref{QKD_with_W-state} we discuss the establishment of a secret conference key where the entanglement resource is a $W$ state. In \autoref{finite-key_analysis} we prove the CKA security in the finite-key scenario (the detailed proof is given in \ref{security_proof}). In \autoref{simulations} we provide simulations of the protocol's secret key rate and compare them with the repeated use of bipartite schemes (further comparisons in \ref{optimizedCKA}). We present our conclusions in \autoref{conclusions}. In \ref{channel_model_computations} we report in detail the calculations of the relevant parameters for an honest implementation of the protocol.

\section{Conference key agreement} \label{CKA-protocol}  
\begin{figure}[!htb]
	\centering
	\includegraphics[width=1\linewidth,keepaspectratio]{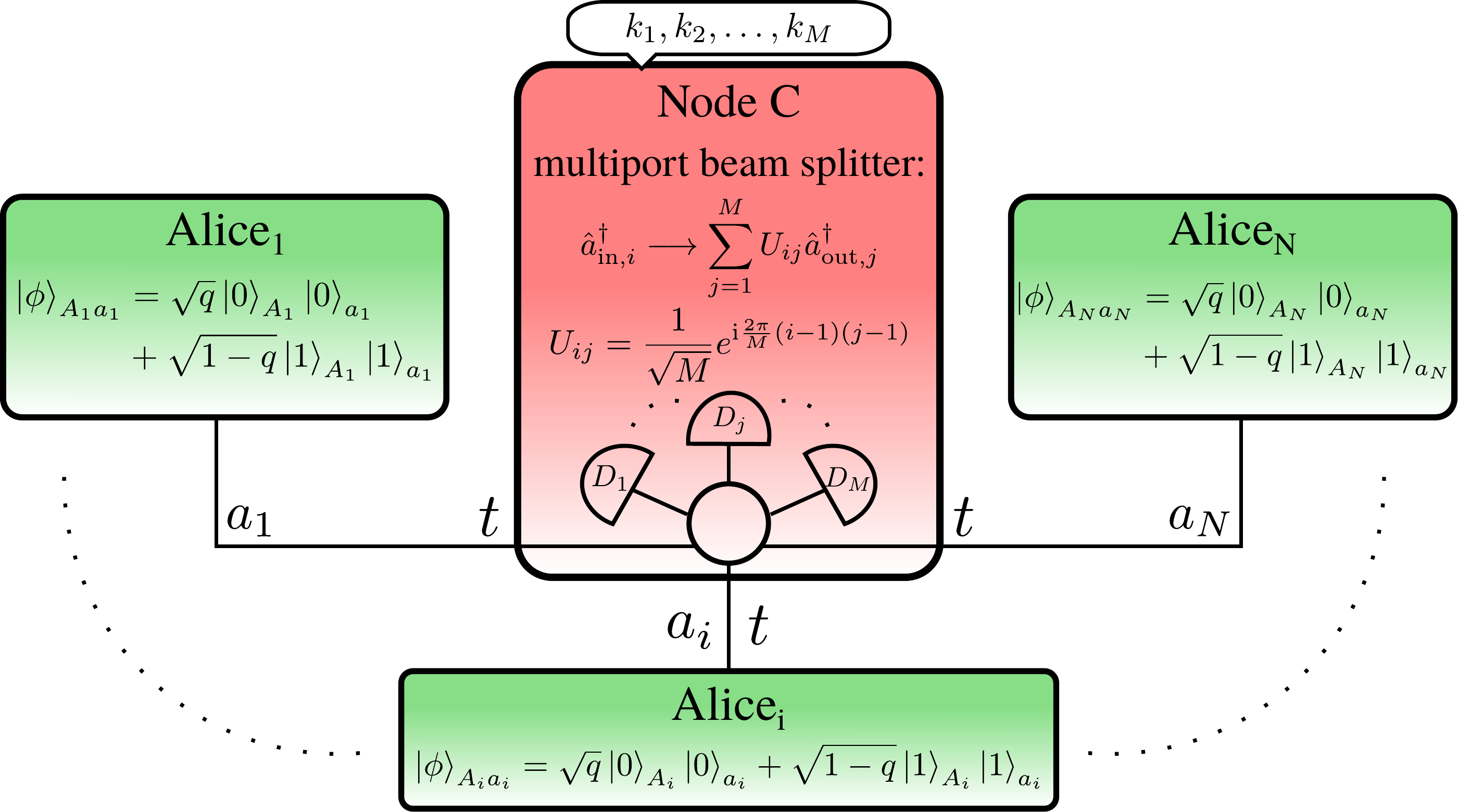}
	\caption{The CKA based on single-photon interference in the untrusted central node and on trusted measurements performed on each party's ($\mathrm{Alice}_i$) qubit.}
	\label{CKA-scheme}
\end{figure}
As anticipated in the introduction, our CKA scheme is an extension of the original bipartite TF-QKD protocol \cite[Protocol 1]{Curty-security-TF} to a scenario with $N$ users who want to establish a secret conference key. The parties distill the secret key by sending optical pulses to an untrusted node and by performing suitable measurements on a qubit they hold. In order to keep the notation symmetric, the $N$ parties involved in the multipartite QKD protocol are named: $\mathrm{Alice}_1$, $\mathrm{Alice}_2$, \dots, $\mathrm{Alice}_N$. The protocol is composed of $L$ rounds, each round is characterized by the following seven steps (see \autoref{CKA-scheme}):
\begin{enumerate}
\item Every party ($\mathrm{Alice}_i$) prepares an optical pulse $a_i$ in an entangled state with a qubit $A_i$, given by:
\begin{equation}
	\ket{\phi}_{A_i a_i} = \sqrt{q} \ket{0}_{A_i}\ket{0}_{a_i}+\sqrt{1-q} \ket{1}_{A_i} \ket{1}_{a_i} \quad \forall\, i\in\{1,2,\dots,N\}\,\,, \label{initial-state}
\end{equation}
where $0\leq q \leq 1$, $\ket{0}_{a_i}$ is the vacuum state, $\ket{1}_{a_i}$ is the single-photon state, and $\{\ket{0}_{A_i},\ket{1}_{A_i} \}$ is the computational basis of qubit $A_i$.
\item Every party sends her optical pulse $a_i$ to the untrusted node via optical channels characterized by transmittance $t$, in a synchronized manner.
\item The central node applies a Bell-multiport beam splitter \cite{first-multiport-bs,multiport-beamsplitter,not-balanced-transmittivities,experimental-tritter,universal-multiport-interferferometers,linear-optics-for-4port-interferometer} with $M$ input and output ports\footnote{We assume that there are at least as many input ports of the beam splitter as parties taking part to the protocol, i.e. $M\geq N$.} to the incoming pulses and features a threshold detector $D_i$ at each output port ($i=1,\dots,M$). The action of the multiport beam splitter is defined by the unitary transformation given in \autoref{CKA-scheme}.
\item The central node announces the measurement outcome $k_i$ for every detector $D_i$, with $k_i=0$ and $k_i=1$ corresponding to a no-click and a click event, respectively. The round gets discarded if $\sum_{i=1}^M k_i \neq 1$, i.e. whenever single-photon interference did not occur in  the central node. The probability that only detector $D_j$ clicked is $p_j$.
\item According to a preshared secret key of $L\cdot h(p_{\mathrm{PE}})$ bits\footnote{Where $h(x)=-x\log_2 x - (1-x) \log_2 (1-x)$ is the binary entropy function.}, the round is classified as a parameter-estimation (PE) round with probability $p_{\mathrm{PE}}$ or as a key-generation (KG) round with probability $1-p_{\mathrm{PE}}$. There are on average $m=M p_j L p_{\mathrm{PE}}$ PE rounds that do not get discarded.
\begin{enumerate}
	\item In case of a PE round, every party measures her qubit in the $Z$-basis and then announces the measurement outcome to compute the frequency: $Q^m_Z=(1+\braket{Z^{\otimes N}}_m)/2$.
	\item In case of a KG round, conditioned on detector $D_j$ clicking, $\mathrm{Alice}_i$ measures her qubit in the basis of the operator $O_{XY}(\varphi_i) = \cos\varphi_i X+\sin\varphi_i Y$ (where $X$ and $Y$ are the Pauli operators), with $\varphi_i=\arg(U_{ij})$ ($U_{ij}$ is given in \autoref{CKA-scheme}). The parties announce $m$ randomly chosen measurement results in order to estimate the quantum bit error rate (QBER) by computing the frequency: $Q^m_{A_1 A_i}= (1-\braket{O_{XY}(\varphi_1) O_{XY}(\varphi_i)}_m)/2$, i.e. the frequency of discordant outcomes.
\end{enumerate}
\item The secret key shared by the $N$ users is extracted from the remaining $n=M p_j L-2m$ raw key bits of the KG rounds.
\item The parties perform classical post-processing and correct their errors to match $\mathrm{Alice}_1$'s raw key.
\end{enumerate}
\textbf{Remarks.} Note that the quantity $Q^m_Z$ is the frequency of the outcome +1 when the parties measure the operator $Z^{\otimes N}$. By making an analogy with the bipartite scenario, one can view $Q^m_Z$ as an estimation of the phase-error rate between $\mathrm{Alice}_1$ and the other $N-1$ parties (when the phase-error rate is defined as in \cite{Curty-security-TF}).\\
Since the Bell-multiport beam splitter redirects each incoming photon with equal probability to each potential output port, the probability of having a click in only one specific detector is the same for all detectors, i.e. $p_j$ reads the same for $j=1,\dots,M$. For this reason, the total probability of having exactly one click in any detector is given by $M p_j$.\bigskip\\
In an honest implementation of the protocol, where the parties' state preparation and the operations of the central node are carried out as described above, the state of the qubits $A_1,\dots,A_N$ from which the parties distill a secret key is approximately a $W$-class state of $N$ qubits \cite{Wstate}, as we show in \autoref{QKD_with_W-state}. Therefore, the protocol here introduced represents an alternative to other multipartite QKD protocols \cite{Epping,Ribeiro,Grasselli,Jo} where the entanglement resource used to generate the key is, instead, a noisy version of the GHZ state of $N$ qubits. Moreover, the $W$-class state used by the CKA is an entangled state which is post-selected after the interference of one single photon at the multiport beam splitter. Thus the resulting key rate scales \textit{linearly} with the transmittance $t$ of one of the quantum channels linking the parties to the central node. This is in contrast to the other mentioned multipartite QKD protocols \cite{Epping,Ribeiro,Grasselli,Jo}, where the distribution of an $N$-qubit GHZ state (e.g. encoded in orthogonal polarizations of a photon) would lead to a key rate which scales with $t^N$ (with $t$ being the transmittance of the link between one party and the node distributing the GHZ state). This makes our CKA much more suited to high-loss scenarios than previously proposed multipartite QKD protocols.

\section{Multipartite QKD with a $W$ state} \label{QKD_with_W-state}
As mentioned at the end of \autoref{CKA-protocol}, the entanglement resource exploited to distill the secret key is a noisy $W$-class state of $N$ qubits \cite{Wstate}, which is post-selected after single-photon interference occurred in the central node. In fact, the optimization of the CKA key rate (\autoref{simulations}) over the parameter $q$ weighting the initial superposition of the qubit-photon state always yields values of $q$ close to 1. This means that the quantum signal sent by the parties is strongly unbalanced towards the vacuum. Thus the events in which one of the detectors clicks are mainly caused by the arrival and detection of one photon. However, because of the balanced superposition generated by the multiport beam splitter, the detected photon could be sent by any party with equal probability. Since the photon is initially entangled to the qubit in state $\ket{1}$, the qubits' state conditioned on the detection is a coherent superposition of states in which one qubit is in state $\ket{1}$ and all the others are in state $\ket{0}$, that is the mentioned $W$-class state. A secret conference key can then be extracted by proper measurements performed on such a state.\\
Let's start by considering the simplistic scenario in which the parties share the $N$-partite $W$ state:
\begin{equation}
	\ket{W}_N = \frac{1}{\sqrt{N}} \left[\ket{00\dots 01}+\ket{00\dots 10} + \dots + \ket{10\dots 00}\right] \label{W-state} \,\,.
\end{equation}
It has been proven \cite{Epping} that the parties cannot extract perfectly correlated outcomes in any set of local measurement bases (for $N\geq 3$). Indeed, the only $N$-qubit state achieving that and yielding uniformly distributed random measurement outcomes is the GHZ state. Nevertheless, the $N$-partite $W$ state can still be used to extract a secret conference key. The key bits are given by the outcomes of the $X$-basis measurements performed by the $N$ parties on their respective qubit. The expected QBER between any two parties is given by $1/2 -1/N$, which amounts to subtracting the fraction $h(1/2-1/N)$ from the secret key rate due to error correction ($h(x)$ is the binary entropy). On the other hand, the eavesdropper's knowledge about the key can be estimated via the phase-error rate $Q_Z$ (as defined in \autoref{CKA-protocol}, more details in \ref{security_proof}), which turns out to be zero on the $W$ state. This is crucial for having a non-zero key rate even when the number of parties is large. The resulting asymptotic key rate, when the parties share an $N$-partite $W$ state, is given by $1-h(1/2-1/N)$.\\
Our CKA is constructed following the same philosophy. The only difference is that the conditional state shared by the parties after the detector's click is not exactly the $W$ state given by (\ref{W-state}), but rather a noisy $W$-class state (the full expression is given in \ref{channel_model_computations}). Indeed, the multiport beam splitter introduces complex phases in the balanced superposition of states shared by the parties, that depend on which detector clicked. For this reason, we require the parties to adjust their KG measurements in the $X,Y$ plane in order to remove such phases and obtain the same QBER they would observe by measuring in the $X$-basis had they shared the standard $W$ state (\ref{W-state}). However, the adjusted KG measurements do not commute with the operations performed in the untrusted node and prevent the CKA to be recast as an MDI prepare-and-measure scheme, opposed to its bipartite version \cite{Curty-security-TF}. Consequently, the multipartite scheme presented here is more challenging to implement than its bipartite counterpart. Nonetheless, the operations that the parties are required to perform seem to be within technological reach \cite{NV-centers-experiment,Wehner-experiment}. In particular, the qubit system could be realized by a nitrogen-vacancy electron spin, whose coherence time has recently reached the order of seconds \cite{long-coherence}. The entanglement between the electron spin and the photon's Fock state would then be generated via selective optical pulses and coherent rotations \cite{NV-centers-experiment}.

\section{Finite-key analysis} \label{finite-key_analysis}
The protocol presented in \autoref{CKA-protocol} can be effectively regarded as an $N$-partite QKD protocol solely characterized by the unknown quantum state $\rho^{M p_j L}_{A_1 A_2 \dots A_N}$, which is the global state of the parties' qubits in all the rounds that were not discarded\footnote{On average, the number of rounds that are not discarded by the CKA is $M p_j L$.}. In this way we allow the eavesdropper, who is in total control of the untrusted node, to perform any kind of operation (coherent attacks) on the whole set of signals sent by the parties in the different rounds. As described above, in each round the parties perform trusted measurements on the state $\rho^{M p_j L}_{A_1 A_2 \dots A_N}$, according to the preshared key they hold. The security of such a multipartite QKD protocol can be proven thanks to the finite-key analysis developed in \cite{Grasselli}. In particular, since $\mathrm{Alice}_1$ (who holds the key to which all the other parties correct their raw key) measures her qubit only in the two mutually unbiased bases $Z$ and $X$, the protocol's security proof follows analogous lines to the one of the $N$-BB84 protocol presented in \cite{Grasselli}. The detailed proof of the CKA security is given in \ref{security_proof}.
\begin{thm}  \label{finite-key}
	The CKA in \autoref{CKA-protocol}, with the optimal 1-way error-correction protocol (which is $\varepsilon_{\mathrm{EC}}$-fully secure and $2(N-1)\varepsilon_{\mathrm{PE}}\,$-robust) and where the secret key generated by two-universal hashing has length
	\begin{eqnarray}
	\fl \ell (N) = &n\, \left[1 - h\left(Q^m_Z + \gamma(n,m,Q^m_Z,\varepsilon_z)\right) - \max_i h\left(Q_{A_1 A_i}^m +\gamma(n,m,Q_{A_1 A_i}^m,\varepsilon_x)\right)\right] 
	- \log_2 \frac{2(N-1)}{\varepsilon_{\mathrm{EC}}}  \nonumber \\
	\fl &- 2\log_2 \frac{1-2(N-1)\varepsilon_{\mathrm{PE}}}{2\,\varepsilon_{\mathrm{PA}}} \,\,, \label{key-length}
	\end{eqnarray}
	is $\varepsilon_{\mathrm{tot}}$-secure with $\varepsilon_{\mathrm{tot}}=2\varepsilon_{\mathrm{PE}} + \varepsilon_{\mathrm{EC}} + \varepsilon_{\mathrm{PA}}$,
	where $\varepsilon_{\mathrm{PE}}$ is defined as:
		\begin{equation}
		\varepsilon_{\mathrm{PE}} \equiv \sqrt{(N-1)\varepsilon_x + \varepsilon_z} 
		\end{equation} 
	and $\gamma(n,m,\Lambda_m,\varepsilon)$ is the positive root of the following equation:
	\begin{equation}
			\ln {n(\Lambda_m+\gamma)+m \Lambda_m \choose m \Lambda_m} + \ln {(n+m)(1-\Lambda_m)-n \gamma \choose m(1-\Lambda_m)} = \ln {n+m \choose m} + \ln \varepsilon  \,\,. \label{gamma-eq}
	\end{equation}
\end{thm}
We remark that the length $L\cdot h(p_{\mathrm{PE}})$ of the preshared key must be subtracted from the secret key length in order to have the net amount of fresh secret key bits.
We also remark that our leakage estimation considers the \textit{worst-case} QBER affecting the parties' raw keys, which is (with high probability) not larger than the QBER observed in appositely designated KG rounds with the appropriate statistical correction. This is in contrast to several other finite-key analyses \cite{ScaraniRenner,Sheridan,Tomamichel,Curty-finiteMDI}, where either the QBER is assumed to be known a priori or its estimation does not account for statistical fluctuations.\\
In the asymptotic regime ($L\rightarrow\infty$), the finite-size effects are not present and the secret key rate ($r=\ell/L$) reads:
\begin{equation}
 r (N) =M p_j \left[1-h(Q_Z) - \max_{i\in\{1,\dots,N\}} h(Q_{A_1 A_i})\right] \,\,,  \label{key-rate}
 \end{equation}
where $Q_Z$ and $Q_{A_1 A_i}$ are the probabilities correspondent to the frequencies defined in \autoref{CKA-protocol}.

\section{Simulations} \label{simulations}
In this Section we provide plots of the secret key rate --number of secret key bits per round-- achieved by the CKA both with finite-key effects (\ref{key-length}) and in the asymptotic regime (\ref{key-rate}), as a function of the loss in one of the channels linking a party to the central node, measured in dB ($-10\log_{10} t$). We assume that the protocol is honestly implemented as described in \autoref{CKA-protocol} and we account for a dark count probability of $p_d=10^{-9}$ in every detector (which can be attained with superconducting nanowire single photon detectors \cite{experiment-Toshiba}) and for a polarization and a phase misalignment between $\mathrm{Alice}_1$ and each other party of 2\%. The relevant error rates and probabilities for this configuration are given in \ref{channel_model_computations}. The plots are optimized over the parameter $q$ of the initial superposition between the two qubit-photon states, unless otherwise stated. The finite-key plots are further optimized over the probability $p_{\mathrm{PE}}$ of performing a PE round and over the security parameters $\varepsilon_x,\varepsilon_z,\varepsilon_{\mathrm{EC}}$ and $\varepsilon_{\mathrm{PA}}$, constrained by a fixed total security parameter of $\varepsilon_{\mathrm{tot}}=10^{-8}$.\\
In order to assess the performance of our CKA with an untrusted node, we consider the situation in which the central node is removed and the $N$ parties are linked by a star network, where the transmittance of the link between any two parties is $t^2$. For this configuration, we consider the conference key rate generated by the following strategy and compare it to our CKA key rate. One selected party performs the best possible bipartite QKD scheme with every other party in the network, i.e. $N-1$ times. Because of the network symmetry, every bipartite secret key has the same length and its asymptotic rate is upper bounded by the Pirandola-Laurenza-Ottaviani-Banchi (PLOB) bound \cite{PLOB} given by: $-\log_2 (1-t^2)$. Then, the selected party encodes the final conference key by using the keys she/he established singularly with each other party. Hence, the conference key length is equal to the bipartite keys' lengths, but the total number of rounds\footnote{By round we mean a set of steps of a given QKD protocol which contains only one transmission of quantum signals (more parties at the same time can transmit a quantum signal).} needed to establish the conference key is given by the number of rounds performed by a pair of parties, multiplied by the number of bipartite schemes ($N-1$). Thus the conference key rate achieved by this strategy is upper bounded by:
\begin{equation}
r_{\mathrm{direct}}(N)= \frac{-\log_2 (1-t^2)}{N-1} \label{PLOB} \,\,.
\end{equation}
We will refer to (\ref{PLOB}) as the \textit{direct-transmission bound}, even though we emphasize that it only upper bounds the achievable conference key rate when the strategy we just described is employed.
Indeed, we do not claim that this strategy yields the highest possible conference key rate for the considered network configuration. In this Section we show that our CKA provides an advantage, in terms of performance, with respect to the above strategy (\ref{PLOB}).

\subsection{Asymptotic regime}
\begin{figure}[!htb]
	\centering
	\includegraphics[width=0.7\linewidth,keepaspectratio]{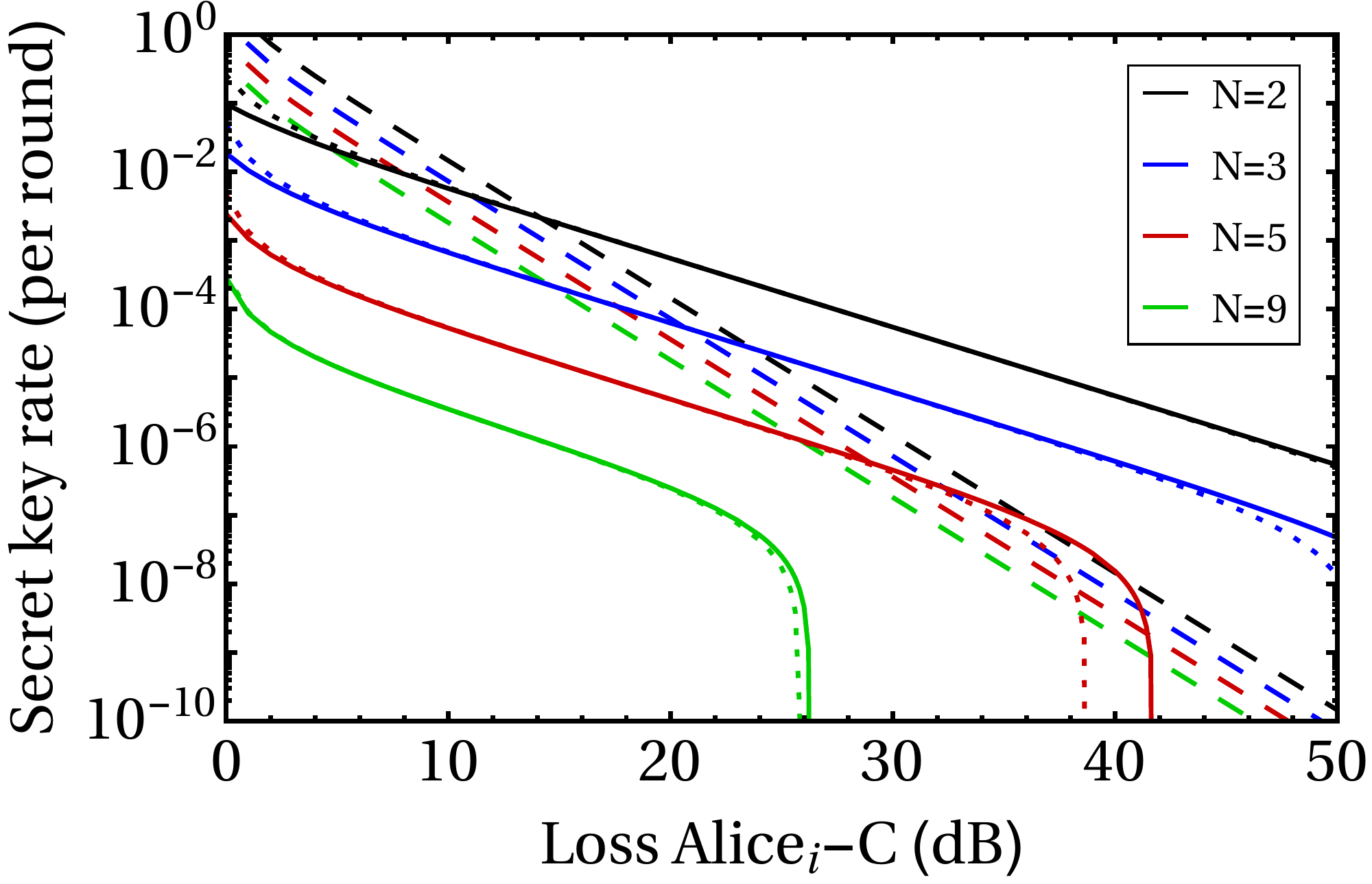}
	\caption{The CKA key rate (Eq.~\ref{key-rate}, solid and dotted lines) and the direct-transmission bound (Eq.~\ref{PLOB}, dashed lines), as a function of the loss in the channel linking one party to the central node, for different number of parties $N=2,3,5$ and $9$ (black, blue, red and green; top to bottom). The CKA key rate overcomes the correspondent direct-transmission bound for increasing losses, as the number of parties increases. For instance, the CKA performed by 5 parties becomes advantageous at distances larger than 150 km (assuming a fiber attenuation of $\alpha=0.2\,\, \mathrm{dB/km}$). We also observe that having more ports in the beam splitter than parties involved in the protocol is advantageous at low losses (dotted lines are above the solid lines) but disadvantageous at high losses, where more ports imply a higher chance of having a dark count.}
	\label{asymptotic_plot}
\end{figure}
In \autoref{asymptotic_plot} we plot the asymptotic key rate of the CKA (Eq.~\ref{key-rate}, solid and dotted lines) as a function of the loss in one of the quantum channels, for different number parties establishing the secret conference key. In particular, the solid lines are obtained by fixing $M=N$, i.e. the number of input (output) ports of the beam splitter is given by the number of parties taking part to the protocol. The dotted lines are instead obtained by fixing the number of ports to $M=10$. Finally, the dashed lines represent the direct-transmission bound (\ref{PLOB}) for the correspondent number of parties.\\
We observe that the CKA key rate can surpass the direct-transmission bound for sufficiently high losses. This is expected since the CKA key rate basically scales linearly with the transmittance $t$ of the quantum channel linking one party to the central node, while the direct-transmission rate scales linearly with the transmittance ($t^2$) of the whole channel linking two parties \cite{PLOB}. We note, however, that the performance advantage of the CKA with respect to the direct-transmission bound decreases for increasing number of parties. This is due to the fact that an increase of the number of parties is more detrimental for the CKA rate, as it severely affects the QBER, than for the direct-transmission bound, where it simply increases the total number of rounds dividing the key length. Moreover, the presence of dark counts in the detectors prevents the CKA from outperforming the direct-transmission bound if the number of parties is too large (see the $N=9$ case in \autoref{asymptotic_plot}). As a matter of fact, for increasing number of parties the key rate optimization yields a lower probability of having a single click in one of the detectors, thus increasing the relative effect of dark counts.\\
From \autoref{asymptotic_plot} we also deduce that performing the CKA with a higher number of ports in the beam splitter (dotted lines, where $M=10$) is advantageous at low losses and disadvantageous at high losses. The advantage of having more output ports is that the probability that two photons arrive at the same detector diminishes (this is an error source in our CKA). However, these errors could only occur if there is a non-negligible probability that two photons arrive at the central node, i.e. when the losses are low. At the same time, the presence of more output ports --and thus detectors-- increases the chances of a dark count. And the negative effect of dark counts on the performance becomes tangible when their probability is comparable to the probability of having a click in a detector, i.e. at high losses.\medskip\\
Another relevant scenario for assessing the CKA performance in comparison to the iteration of bipartite protocols could be the following. The parties are given the same CKA experimental setup but they are now allowed to use it in pairs (or larger subgroups) in consecutive runs, effectively performing the original TF-QKD protocol \cite[Protocol 1]{Curty-security-TF} between one selected party and every other party. The different established keys are then used to encode the final conference key, similarly to the direct-transmission scenario. This strategy can then be compared to the case where the parties choose to use the CKA setup all at once, thus performing a truly multipartite QKD scheme. A detailed analysis of this comparison in the asymptotic regime is given in \ref{optimizedCKA}. It turns out that, depending on the loss and on the state preparation, it is still advantageous to perform a multipartite protocol instead of iteratively executing bipartite protocols, on the CKA experimental setup.

\subsection{Finite-key effects}
\begin{figure}[!htb]
	\centering
	\begin{subfigure}[t]{.5\textwidth}
		\centering
		\includegraphics[width=1\linewidth,keepaspectratio]{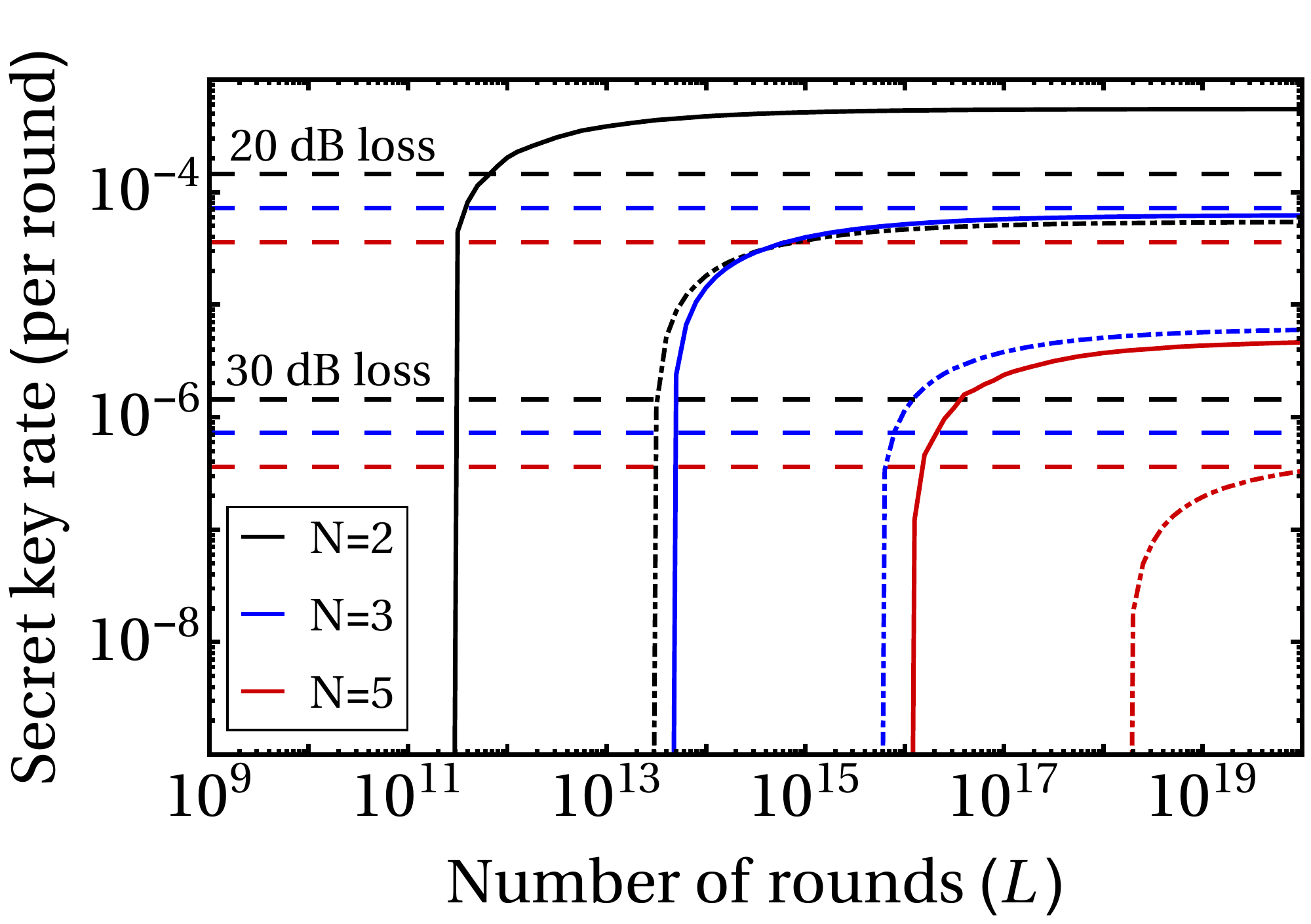}
		\caption{Finite-key conference rate (Eq.~\ref{key-length} over $L$) as a function of the number of rounds $L$, for fixed losses of 20 dB (solid lines) and 30 dB (dot-dashed lines), and different number of parties: $N=2,3$ and $5$ (black, blue and red; top to bottom). We observe that the rates quickly achieve their asymptotic value once the number of non-discarded rounds is enough to get a non-zero key. The CKA key rates overcome the direct-transmission bound (dashed lines) even in the finite-key scenario.}
		\label{finite-key_plot}
	\end{subfigure}%
	\begin{subfigure}[t]{.5\textwidth}
		\centering
		\includegraphics[width=1\linewidth,keepaspectratio]{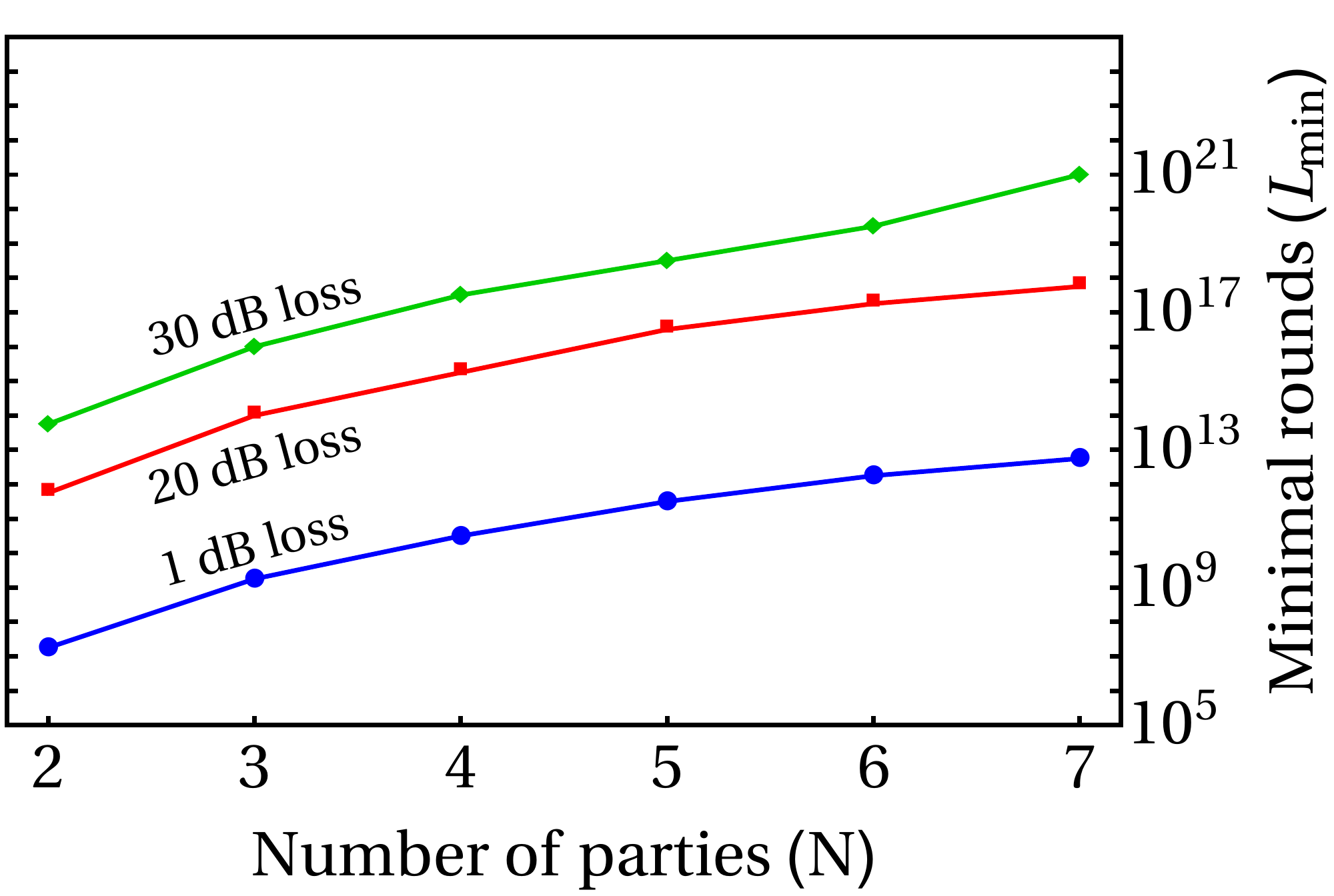}
		\caption{Minimum number of rounds such that the finite-key rate (Eq.~\ref{key-length} over $L$) is at least 10\% of its asymptotic value, as a function of the number of parties and for fixed losses (1 dB blue circles, 20 dB red squares and 40 dB green diamonds). We notice that increasing the number of parties and/or the losses is more detrimental for the finite-key rate than for the asymptotic one, due to an increase of the fraction of discarded rounds and thus of the statistical fluctuations.}
		\label{minimum_rounds}
	\end{subfigure}
\caption{Here we study the finite-key effects on our CKA. The number of ports in the beam splitter is given by the number of parties taking part to the protocol: $M=N$.}
\label{finite-effects}
\end{figure}
In \autoref{finite-key_plot} we plot the finite-key conference rate (Eq.~\ref{key-length} divided by $L$) as a function of the number of rounds $L$, for different fixed values of the loss (20dB and 30dB, solid and dot-dashed lines) and different number of parties. We stress the fact that we normalize the key length to the \textit{total} number of rounds ($L$), i.e. we also take into account the rounds that get discarded due to double-clicks or no click in the detectors. The horizontal dashed lines correspond to the value of the direct-transmission bound (\ref{PLOB}) for the various combinations of losses and number of parties. We observe that the number of rounds leading to a non-zero key rate is in general higher than other multipartite schemes (see for example \cite{Grasselli}). This is caused by the fact that the CKA devised here relies on single-photon interference events, which are only a fraction of all the events occurring in an experiment run. A considerable amount of rounds gets thus discarded, but still contributes to the rounds' count. Nevertheless, the number of rounds needed for a non-zero key rate is comparable to other bipartite TF-QKD protocols \cite{tail-ineq,finite-key-TF}. On the other hand, the advantage of relying on single-photon interference in a multipartite scenario is the excellent scaling of the protocol's key rate with respect to losses, which allows it to overcome the asymptotic direct-transmission bound (dashed lines) even with a finite number of rounds.\medskip\\
In \autoref{minimum_rounds} we instead plot the minimum number of rounds ($L_{\mathrm{min}}$) such that the finite-key rate ($\ell/L$) does not decrease more than 90\% with respect to its asymptotic value $r$ (\ref{key-rate}), i.e.: $\ell(L_{\mathrm{min}})/L_{\mathrm{min}} \geq r/10$. The threshold $L_{\mathrm{min}}$ is plotted as a function of the number of parties ($N$) and for fixed values of the loss. We observe that $L_{\mathrm{min}}$ increases both with the number of parties and with the loss. The reason is that, in both cases, the fraction of the total number of rounds that gets discarded increases. This has a negative effect both on the asymptotic rate and on the finite-key rate, however the effect on the latter is greater, thus requiring a larger number of rounds $L_{\mathrm{min}}$ to maintain the finite-key rate within 90\% range of the asymptotic one. Indeed, a larger fraction of discarded rounds decreases the prefactor $M p_j$ in both the finite- and the asymptotic-key rates, but it additionally decreases the number of rounds used for PE in the finite-key regime. This causes larger statistical fluctuations and thus a smaller finite-key rate.

\section{Conclusions} \label{conclusions}
In this work we introduced a new multipartite QKD protocol that exploits for the first time the correlations derived from an $N$-partite $W$ state \cite{Wstate} to establish a secret conference key among the $N$ users. In an honest implementation of the protocol, the $W$ state is post-selected thanks to the interference of a \textit{single} photon in a central node, extending the idea of the bipartite Twin-Field QKD protocol devised in \cite{Curty-security-TF} to the multipartite scenario. Hence the resulting key rate scales linearly with the transmittance of one of the quantum channels linking the parties to the central node, making the protocol particularly suited for conference keys established in high-loss scenarios.\\
We prove the protocol's security in the finite-key regime by considering the most adversarial situation possible, i.e. coherent attacks are allowed by the eavesdropper. In order to achieve this, we rely on previous results on the finite-key security of multipartite QKD schemes derived in \cite{Grasselli} and employ the entropic uncertainty relation \cite{uncert-rel}.\\
We provide simulations of the conference key rate both in the finite- and in the asymptotic-key regime. We compare the performance of our conference key agreement (CKA) to that achieved by performing bipartite QKD schemes between one party and each of the others and then using the established keys to encode the conference key. In particular, we analyze the cases where the bipartite schemes are performed with the same setup used for the CKA (in \ref{optimizedCKA}) and in the direct-transmission scenario (i.e. the central node is removed and the optimal bipartite QKD scheme is performed). We show that, in both cases, the execution of a truly multipartite scheme could be advantageous even when finite-key effects are accounted for.\\
Although the feasibility of the proposed CKA requires further investigation, with this work we demonstrate that, in principle, multipartite QKD does not necessarily need a GHZ-class state as its entanglement resource and that it can be implemented even in high-loss scenarios.

\ack
We thank Marcos Curty, Daniel Miller and Hua-Lei Yin for helpful discussions.
This project has received funding from the European Union’s Horizon 2020 research and innovation programme under the Marie Skłodowska-Curie grant agreement No 675662.

	\appendix
	\section{Security proof} \label{security_proof}
	In order to prove the security of our CKA in the finite-key scenario, we start from the general security statement given in \cite[Theorem 1]{Grasselli}. The resulting secret key length is thus determined by the amount of information the eavesdropper has about the secret key and by the information the parties leak during the classical post-processing. The former is quantified by the min-entropy $H_{\mathrm{min}}^{\varepsilon} (\rho^n_{X E} | E)$, where $\rho^n_{X E}$ is the classical-quantum state of $\mathrm{Alice}_1$'s raw key and the eavesdropper's quantum system $E$ which is partially correlated to it. Since the eavesdropper's system is unknown, one cannot directly compute the mentioned min-entropy. Nevertheless, it can be bounded by means of the uncertainty relation \cite{uncert-rel} as follows:
	\begin{eqnarray}
		H_{\mathrm{min}}^{\varepsilon} (\rho^n_{X E} | E) \geq n - H_{\mathrm{max}}^{\varepsilon} (\rho^n_{Z_1\dots Z_N} | Z_2 \dots Z_N )  \,\,,
	\end{eqnarray}
	where the max-entropy on the r.h.s. quantifies the uncertainty of $\mathrm{Alice}_1$'s $Z$-measurement results when the $Z$-outcomes of the remaining $N-1$ parties are known, if all parties would measure $Z$ in the $n$ rounds yielding the raw-key. The max-entropy can be upper bounded via the phase-error rate $Q^n_Z$ --as defined in \autoref{CKA-protocol}-- of the $n$ raw-key rounds. We get:
	\begin{eqnarray}
	H_{\mathrm{min}}^{\varepsilon} (\rho^n_{X E} | E) \geq n - n\,h (Q^n_Z) \,\,,
	\end{eqnarray}
	where $h(\cdot)$ is the binary entropy function: $h(x)=-x\log_2(x) -(1-x)\log_2(1-x)$. Finally, since the parties do not directly observe the phase-error rate $Q^n_Z$ of the $n$ rounds producing the raw key, this can be inferred through the theory of random sampling without replacement. In particular, the phase-error rate of the raw key ($Q^n_Z$) can be upper bounded with high probability, once the observed phase-error rate ($Q^m_Z$) is known. For this, we make use of the following tail inequality \cite[Lemma 1]{tail-ineq} which features a tighter bound with respect to the Serfling inequality.
	\begin{lemma} \label{tail_inequality-lemma}
		{\normalfont \cite{tail-ineq}.} Let $\mathcal{X}_{n+m}$ be a random binary string of $n+m$ bits, $\mathcal{X}_m$ be a random sample (without replacement) of m entries from the string $\mathcal{X}_{n+m}$ and $\mathcal{X}_n$ be the remaining bit string. Upon calling $\Lambda_m$ and $\Lambda_n$ the frequencies of bit value 1 in string $\mathcal{X}_m$ and $\mathcal{X}_n$, respectively, for any $\varepsilon>0$ it holds:
		\begin{equation}
		\mathrm{Pr} [\Lambda_n \leq \Lambda_m + \gamma (n,m,\Lambda_m,\varepsilon)] > 1- \varepsilon  \label{tail_inequality} \,\,,
		\end{equation}
		where $\gamma(n,m,\Lambda_m,\varepsilon)$ is the positive root of the following equation:
		\begin{equation}
		\ln {n(\Lambda_m+\gamma)+m \Lambda_m \choose m \Lambda_m} + \ln {(n+m)(1-\Lambda_m)-n \gamma \choose m(1-\Lambda_m)} = \ln {n+m \choose m} + \ln \varepsilon  \,\,. \label{gamma-eq-appendix}
		\end{equation}
	\end{lemma}
	By applying \autoref{tail_inequality-lemma} to the case of $Q^n_Z$, we can finally bound the eavesdropper knowledge about the secret key as follows:
	\begin{eqnarray}
	H_{\mathrm{min}}^{\varepsilon} (\rho^n_{X E} | E) \geq n - n\, h\left(Q^m_Z + \gamma(n,m,Q^m_Z,\varepsilon_z)\right) \,\,.
	\end{eqnarray}
	The remaining part of the secret key length that needs to be estimated is the leaked information during classical post-processing.
	This is readily estimated through the observed QBER between $\mathrm{Alice}_1$ and any other party, by means of \cite[Theorem 2]{Grasselli}.\\
	Putting these considerations together, we obtain the security statement given in \autoref{finite-key}.
	
	\section{Channel model}  \label{channel_model_computations}
	In this Section we compute the QBER ($Q_{A_1 A_k}$), the phase-error rate ($Q_Z$) and the probability that a given detector clicked ($p_j$), assuming that the protocol is implemented as described in \autoref{CKA-protocol}. We also account for a dark count probability $p_d$ in each detector and we consider the specific scenario in which there are a polarization and a phase misalignment of angles $\theta$ and $\phi$, respectively, between $\mathrm{Alice}_1$ and each other party. In the simulations of \autoref{simulations} we set: $p_d=10^{-9}$ and $\theta=\phi=\arcsin\sqrt{0.02}$. For simplicity, we assume that the input signals of the $N$ parties enter the first $N$ ports of the $M$-port beam splitter. Nevertheless, the results in terms of achieved key rate are independent of which input ports are used, thanks to the balanced redistribution of the input photons to the output ports of the considered Bell-multiport beam splitter (see \autoref{CKA-scheme}). We remark that the expressions derived here together with the asymptotic key rate given in (\ref{key-rate}) reproduce those of the original TF-QKD protocol \cite[Protocol 1]{Curty-security-TF} in the case of two parties ($N=2$) with a balanced 2-port beam splitter ($M=2$).\medskip\\
	We first derive the QBER, the phase-error rate and the probability $p_j$ assuming no dark counts in the detectors, i.e. every click is caused by the arrival of one or more photons. In the last Subsection we use the derived expressions to obtain analogous quantities, with the assumption that every detector has a probability $p_d$ of clicking conditioned on no photon arriving.
	
	\subsection{Qubits' state conditioned on one click}
	According to the protocol, the global state of the parties' qubits and signals, before sending the signals to the central node, reads:
	\begin{equation}
	\ket{\Phi_1} = \bigotimes_{k=1}^N \ket{\phi}_{A_k a_k} = \bigotimes_{k=1}^N \left(\sqrt{q} \ket{0}_{A_k}\ket{0}_{a_k}+\sqrt{1-q} e^{\mathrm{i}\phi_k}\ket{1}_{A_k} a_k^\dag\ket{0}_{a_k}\right) \,\,,  \label{Phi1}
	\end{equation}
	where the phase mismatch $\phi_k$ is defined as zero if $k=1$ and as $\phi$ if $k\neq 1$, which means that every other party has the same phase mismatch with respect to $\mathrm{Alice}_1$. The signals $a_k$ are then sent to the central node through lossy optical channels, which are modeled as beam splitters with transmittance $t$. The global state after the transmission of the signals to the untrusted relay reads:
	\begin{eqnarray}
	\ket{\Phi_2} &= \bigotimes_{k=1}^N \left[\sqrt{q} \ket{0}_{A_k}\ket{0}+\sqrt{1-q}e^{\mathrm{i}\phi_k} \ket{1}_{A_k} (\sqrt{t}\, a_k^\dag +\sqrt{1-t} \, l_k^\dag )\ket{0}\right] \nonumber\\
	&= \sum_{g(\vec{b})=0}^{2^N-1} q^{\frac{N-|\vec{b}|}{2}} (1-q)^{\frac{|\vec{b}|}{2}} \ket{\vec{b}}_{A_1\dots A_N} \otimes_{k=1}^{N} e^{\mathrm{i}b_k \phi_k}(\sqrt{t}\, a_k^\dag +\sqrt{1-t} \, l_k^\dag )^{b_k} \ket{0} \,\,,   \label{Phi2}
	\end{eqnarray}
	where $l_k^\dagger$ is the creation operator of the lost photon in channel $k$, $\vec{b}$ is a $N$-bit vector that runs from 0 to $2^N-1$ in binary notation (covering all the possible combinations of qubit states) and $|\vec{b}|$ is the Hamming weight of vector $\vec{b}$. From now on, we denote as $g(\cdot)$ the bijective function that takes as input a binary vector and outputs the correspondent decimal number.\\
	We assume now that the polarization of the photons sent by $\mathrm{Alice}_2,\dots\mathrm{Alice}_N$ is rotated by an angle $\theta$ with respect to $\mathrm{Alice}_1$'s signal:
	\begin{eqnarray}
	\fl \ket{\Phi_3} = \sum_{g(\vec{b})=0}^{2^N-1} q^{\frac{N-|\vec{b}|}{2}} (1-q)^{\frac{|\vec{b}|}{2}} \ket{\vec{b}}_{A_1\dots A_N} \otimes_{k=1}^{N} e^{\mathrm{i}b_k \phi_k}(\sqrt{t}\, \cos\theta_k a_{k,\mathrm{P}}^\dag - \sqrt{t}\, \sin\theta_k a_{k,\mathrm{P}_\bot}^\dag +\sqrt{1-t} \, l_k^\dag )^{b_k} \ket{0} \,,   \label{Phi3}
	\end{eqnarray}
	where $\theta_k$ is defined as zero if $k=1$ and as $\theta$ if $k \neq 1$, while the subscripts ${}_{\mathrm{P}}$ and ${}_{\mathrm{P}_\bot}$ indicate the polarization of $\mathrm{Alice}_1$'s signal and its orthogonal direction, respectively.\\
	Finally, the global state after the application of the Bell-multiport beam splitter on the incoming signals (its action on the incoming creation operators is reported in \autoref{CKA-scheme}) is:
	\begin{eqnarray}
	\fl \ket{\Phi_4} = \sum_{g(\vec{b})=0}^{2^N-1} q^{\frac{N-|\vec{b}|}{2}} (1-q)^{\frac{|\vec{b}|}{2}} \ket{\vec{b}}_{A_1\dots A_N} \nonumber \\
	\fl \otimes \prod_{k=1}^{N} e^{\mathrm{i}b_k \phi_k} (\sqrt{t}\, \cos\theta_k \textstyle\sum_{j=1}^M U_{kj} \sigma_{j,\mathrm{P}}^\dag - \sqrt{t}\, \sin\theta_k \sum_{j=1}^M U_{kj} \sigma_{j,\mathrm{P}_\bot}^\dag +\sqrt{1-t} \, l_k^\dag )^{b_k} \ket{0} \,, \label{Phi4}
	\end{eqnarray}
	where $\sigma_{j,\mathrm{P}}^\dag$ and $ \sigma_{j,\mathrm{P}_\bot}^\dag$ are the creation operators of the output signals in the two orthogonal polarizations and $U_{kj}$ is reported in \autoref{CKA-scheme}. At this point, every output signal is measured in the respective threshold detector. Since the detectors do not distinguish the polarization of the output signals, we will use the subscript ${}_{\sigma_j}$ to indicate the combined Hilbert space of the signals exiting port $j$, when there is no ambiguity.\\
	We are now ready to compute the conditional state of the qubits $A_1,\dots,A_N$ when only detector $D_j$ clicked:
	\begin{eqnarray}
	\fl&p_j \rho_{A_1\dots A_N}^j = \Tr_{\stackrel[l_1,\dots,l_N]{\sigma_1,\dots,\sigma_M}{}}\left[(\mathrm{id}_{\sigma_j}-P_{\ket{0}_{\sigma_j}})\otimes_{i\neq j} P_{\ket{0}_{\sigma_i}} \ket{\Phi_4}\bra{\Phi_4}(\mathrm{id}_{\sigma_j}-P_{\ket{0}_{\sigma_j}})\otimes_{i\neq j} P_{\ket{0}_{\sigma_i}} \right] \label{rho-j-comput1}
	\end{eqnarray}
	where $p_j$ is the probability that only detector $D_j$ clicked, $\rho_{A_1\dots A_N}^j$ is the normalized conditional state of the qubits and $P_{\ket{0}_{\sigma_j}}$ is the projector on the vacuum state of output signal $j$. In order to compute (\ref{rho-j-comput1}), we start by calculating the following quantity:
	\begin{eqnarray}
	\fl &(\mathrm{id}_{\sigma_j}-P_{\ket{0}_{\sigma_j}})\otimes_{i\neq j} P_{\ket{0}_{\sigma_i}} \ket{\Phi_4} = \sum_{g(\vec{b})=1}^{2^N-1} q^{\frac{N-|\vec{b}|}{2}} (1-q)^{\frac{|\vec{b}|}{2}} \ket{\vec{b}}_{A_1\dots A_N} \nonumber\\
	\fl &\otimes (\mathrm{id}_{\sigma_j}-P_{\ket{0}_{\sigma_j}})\otimes_{i\neq j} P_{\ket{0}_{\sigma_i}} \prod_{k=1}^{N} e^{\mathrm{i}b_k \phi_k} \left[\sqrt{t}U_{kj} \, \left(\cos\theta_k \sigma_{j,\mathrm{P}}^\dag - \sin\theta_k \sigma_{j,\mathrm{P}_\bot}^\dag\right) +\sqrt{1-t} \, l_k^\dag \right]^{b_k} \ket{0} \nonumber\\
	\fl & \equiv \sum_{g(\vec{b})=1}^{2^N-1} q^{\frac{N-|\vec{b}|}{2}} (1-q)^{\frac{|\vec{b}|}{2}} \ket{\vec{b}}_{A_1\dots A_N} \otimes (\mathrm{id}_{\sigma_j}-P_{\ket{0}_{\sigma_j}})\otimes_{i\neq j} P_{\ket{0}_{\sigma_i}} \ket{\psi}_{\sigma_j,l}\label{rho-j-comput2}
	\end{eqnarray}
	where the effect of the projectors is to select the outcome signal $\sigma_j$ and to remove the case $g(\vec{b})=0$, since it would correspond to a vacuum state for the outcome signal $\sigma_j$. We now focus on rewriting the following term:
	\begin{eqnarray}
	\fl &\ket{\psi}_{\sigma_j,l} = \prod_{k=1}^{N} e^{\mathrm{i}b_k \phi_k} \left[\sqrt{t}U_{kj} \, \left(\cos\theta_k \sigma_{j,\mathrm{P}}^\dag - \sin\theta_k \sigma_{j,\mathrm{P}_\bot}^\dag\right) +\sqrt{1-t} \, l_k^\dag \right]^{b_k} \ket{0} = \nonumber\\
	\fl & e^{\mathrm{i}(|\vec{b}|-b_1)\phi} \sum_{\stackrel[\vec{d}\wedge\vec{b}=\vec{d}]{g(\vec{d})=0 \mathrm{\,s.t.}}{}}^{2^N-1} \prod_{k=1}^{N} \left[\sqrt{t}\, U_{kj}\left(\cos\theta_k \sigma_{j,\mathrm{P}}^\dag - \sin\theta_k \sigma_{j,\mathrm{P}_\bot}^\dag\right)\right]^{d_k} \left(\sqrt{1-t} \, l_k^\dag\right)^{(\vec{b}\oplus\vec{d})_k} \ket{0}  \label{rho-j-comput3} \\
	\fl &= e^{\mathrm{i}(|\vec{b}|-b_1)\phi} \sum_{\stackrel[\vec{d}\wedge\vec{b}=\vec{d}]{g(\vec{d})=0 \mathrm{\,s.t.}}{}}^{2^N-1} e^{\mathrm{i}\frac{2\pi}{M}(j-1)\sum_{k=1}^{N} d_k (k-1)} \left(\sqrt{\frac{t}{M}}\right)^{|\vec{d}|} (\sqrt{1-t})^{|\vec{b}\oplus\vec{d}|} \nonumber\\
	\fl &\hspace{13ex}\times\prod_{k=1}^{N} \left(\cos\theta_k \sigma_{j,\mathrm{P}}^\dag - \sin\theta_k \sigma_{j,\mathrm{P}_\bot}^\dag\right)^{d_k} \ket{0} \otimes \ket{\vec{b}\oplus \vec{d}}_{l_1,\dots,l_N} \,\,, \label{rho-j-comput4}
	\end{eqnarray}
	where we expanded the product in the first line of (\ref{rho-j-comput3}) by introducing a sum over the binary vector $\vec{d}$. The sum runs over all the $N$-bit vectors $\vec{d}$ for which $d_k=0$ whenever $b_k=0$, for all $k$ --the condition $d_k \wedge b_k=d_k\,\, \forall\,k$. This is to make sure that the $k$-th factor in the first line does not contribute to the expanded product in the second line whenever $b_k=0$. The remaining bits of $\vec{d}$ that are not affected by the mentioned condition, can be either 1 or 0. If $d_k=1$ we intend that, for this particular term in the sum, the contribution of the $k$-th factor in the first line of (\ref{rho-j-comput3}) is given by its first addend ($\sqrt{t}\, U_{kj}(\dots)$). While if $d_k=0$ and $b_k=1$, we mean that the contribution is coming from the second addend ($\sqrt{1-t} \, l_k^\dag$). The exponents in the second line of (\ref{rho-j-comput3}) are chosen according to these rules. Finally, (\ref{rho-j-comput4}) is obtained by using the definition of $U_{kj}$ from \autoref{CKA-scheme} and by applying the creation operators on the vacuum. We now expand the remaining product in (\ref{rho-j-comput4}) with the same technique and obtain the following expression:
	\begin{eqnarray}
	\fl &\ket{\psi}_{\sigma_j,l} = e^{\mathrm{i}(|\vec{b}|-b_1)\phi} \sum_{\stackrel[\vec{d}\wedge\vec{b}=\vec{d}]{g(\vec{d})=0 \mathrm{\,s.t.}}{}}^{2^N-1} e^{\mathrm{i}\frac{2\pi}{M}(j-1)\sum_{k=1}^{N} d_k (k-1)} \left(\sqrt{\frac{t}{M}}\right)^{|\vec{d}|} (\sqrt{1-t})^{|\vec{b}\oplus\vec{d}|} \nonumber\\
	\fl &\times\sum_{\stackrel[\vec{f}\wedge\vec{d}=\vec{f}]{g(\vec{f})=0 \mathrm{\,s.t.}}{}}^{2^N -1} \prod_{k=1}^N (\cos\theta_k)^{f_k} (-\sin\theta_k)^{(\vec{d}\oplus \vec{f})_k} \sqrt{|\vec{f}|!} \sqrt{|\vec{d}\oplus \vec{f}|!} \,\,\ket{|\vec{f}|}_{\sigma_j, \mathrm{P}} \otimes \ket{|\vec{d}\oplus \vec{f}|}_{\sigma_j, \mathrm{P}_\bot}\otimes \ket{\vec{b}\oplus \vec{d}}_{l_1,\dots,l_N} \nonumber\\
	\fl &= e^{\mathrm{i}(|\vec{b}|-b_1)\phi} \sum_{\stackrel[\vec{d}\wedge\vec{b}=\vec{d}]{g(\vec{d})=0 \mathrm{\,s.t.}}{}}^{2^N-1} e^{\mathrm{i}\frac{2\pi}{M}(j-1)\sum_{k=1}^{N} d_k (k-1)} \left(\sqrt{\frac{t}{M}}\right)^{|\vec{d}|} (\sqrt{1-t})^{|\vec{b}\oplus\vec{d}|} \nonumber\\
	\fl &\times\sum_{\stackrel[\vec{f}\wedge\vec{d}=\vec{f}]{g(\vec{f})=0 \mathrm{\,s.t.}}{}}^{2^N -1} (\cos\theta)^{|\vec{f}|-f_1} (-\sin\theta)^{|\vec{d}\oplus \vec{f}|} \delta _{(\vec{d}\oplus \vec{f})_1,0}\sqrt{|\vec{f}|!} \sqrt{|\vec{d}\oplus \vec{f}|!} \,\,\ket{|\vec{f}|}_{\sigma_j, \mathrm{P}} \otimes \ket{|\vec{d}\oplus \vec{f}|}_{\sigma_j, \mathrm{P}_\bot}\otimes \ket{\vec{b}\oplus \vec{d}}_{l_1,\dots,l_N} \label{rho-j-comput5}
	\end{eqnarray}
	We now substitute (\ref{rho-j-comput5}) back into (\ref{rho-j-comput2}) and note that the effect of the projectors $(\mathrm{id}_{\sigma_j}-P_{\ket{0}_{\sigma_j}})\otimes_{i\neq j} P_{\ket{0}_{\sigma_i}}$ is to remove the case $g(\vec{d})=0$ from (\ref{rho-j-comput5}). Hence we get:
	\begin{eqnarray}
	\fl &(\mathrm{id}_{\sigma_j}-P_{\ket{0}_{\sigma_j}})\otimes_{i\neq j} P_{\ket{0}_{\sigma_i}} \ket{\Phi_4} =	\nonumber\\
	\fl &\sum_{g(\vec{b})=1}^{2^N-1} q^{\frac{N-|\vec{b}|}{2}} (1-q)^{\frac{|\vec{b}|}{2}} e^{\mathrm{i}(|\vec{b}|-b_1)\phi} \ket{\vec{b}}_{A_1\dots A_N} \otimes \sum_{\stackrel[\vec{d}\wedge\vec{b}=\vec{d}]{g(\vec{d})=1 \mathrm{\,s.t.}}{}}^{2^N-1} e^{\mathrm{i}\frac{2\pi}{M}(j-1)\sum_{k=1}^{N} d_k (k-1)} \left(\sqrt{\frac{t}{M}}\right)^{|\vec{d}|} (\sqrt{1-t})^{|\vec{b}\oplus\vec{d}|} \nonumber\\
	\fl & \times\sum_{\stackrel[\vec{f}\wedge\vec{d}=\vec{f}]{g(\vec{f})=0 : f_1=d_1}{}}^{2^N -1} (\cos\theta)^{|\vec{f}|-f_1} (-\sin\theta)^{|\vec{d}\oplus \vec{f}|} \sqrt{|\vec{f}|!} \sqrt{|\vec{d}\oplus \vec{f}|!} \,\,\ket{|\vec{f}|}_{\sigma_j, \mathrm{P}} \otimes \ket{|\vec{d}\oplus \vec{f}|}_{\sigma_j, \mathrm{P}_\bot}\otimes \ket{\vec{b}\oplus \vec{d}}_{l_1,\dots,l_N} \,\,. \label{rho-j-comput6}
	\end{eqnarray}
	By substituting (\ref{rho-j-comput6}) into (\ref{rho-j-comput1}) we finally get the state of the qubits conditioned on $D_j$ clicking:
	\begin{eqnarray}
	\fl &p_j \rho_{A_1\dots A_N}^j = \sum_{g(\vec{b}),g(\vec{b'})=1}^{2^N-1} q^{N-\frac{|\vec{b}|+|\vec{b'}|}{2}} (1-q)^{\frac{|\vec{b}|+|\vec{b'}|}{2}}  e^{\mathrm{i}[|\vec{b}|-|\vec{b'}|-(b_1-b_1')]\phi} \ket{\vec{b}}\bra{\vec{b'}} \nonumber\\
	\fl &\times \sum_{\stackrel[\vec{d'}\wedge\vec{b'}=\vec{d'}]{g(\vec{d}),g(\vec{d'})=1\,:\,\vec{d}\wedge\vec{b}=\vec{d}}{}}^{2^N-1} e^{\mathrm{i}\frac{2\pi}{M}(j-1)\sum_{k=1}^{N} (k-1)(d_k-d_k')} \left(\sqrt{\frac{t}{M}}\right)^{|\vec{d}|+|\vec{d'}|} (\sqrt{1-t})^{|\vec{b}\oplus\vec{d}|+|\vec{b'}\oplus\vec{d'}|} \nonumber\\
	\fl &\times\sum_{\stackrel[\vec{f}\wedge\vec{d}=\vec{f}]{g(\vec{f})=0 : f_1=d_1}{}}^{2^N -1} \sum_{\stackrel[\vec{f'}\wedge\vec{d'}=\vec{f'}]{g(\vec{f'})=0 : f_1'=d_1'}{}}^{2^N -1} (\cos\theta)^{|\vec{f}|+|\vec{f'}|-f_1-f_1'} (-\sin\theta)^{|\vec{d}\oplus \vec{f}|+|\vec{d'}\oplus \vec{f'}|}  \nonumber\\
	\fl&\times \sqrt{|\vec{f}|!\,|\vec{f'}|!\,|\vec{d}\oplus \vec{f}|!\,|\vec{d'}\oplus \vec{f'}|!} \,\, \delta_{|\vec{f}|,|\vec{f'}|}\,\, \delta_{|\vec{d}\oplus \vec{f}|,|\vec{d'}\oplus \vec{f'}|} \,\,\delta_{\vec{b}\oplus\vec{d},\vec{b'}\oplus\vec{d'}} \,\,. \label{rho-j-comput7} 
	\end{eqnarray}
	We use the Kronecker deltas to reduce the sums over $\vec{d}, \vec{d'}, \vec{f}$ and $\vec{f'}$. The third delta fixes the value of $\vec{d'}$: $\vec{d'}=\vec{b}\oplus\vec{b'}\oplus\vec{d}$. The fixed value of $\vec{d'}$ combined with the other constraints on this vector imply additional constraints on $\vec{d}$. In particular, $\vec{d'}\neq 0$ implies $\vec{d} \neq \vec{b}\oplus \vec{b'}$ while $\vec{d'}\wedge\vec{b'}=\vec{d'}$ implies $\vec{b'} \wedge (\vec{b}\oplus \vec{d}) = \vec{b} \oplus \vec{d}$. Finally the first two deltas imply $|\vec{d}|=|\vec{d'}|$, which combined with the third delta yields $|\vec{b}|=|\vec{b'}|$. Putting everything together allows to simplify (\ref{rho-j-comput7}) as follows:
	\begin{eqnarray}
	\fl &p_j \rho_{A_1\dots A_N}^j = \sum_{\stackrel[|\vec{b'}|=|\vec{b}|]{g(\vec{b}),g(\vec{b'})=1}{}}^{2^N-1} q^{N-|\vec{b}|} (1-q)^{|\vec{b}|}  e^{\mathrm{i}[(b_1'-b_1)]\phi} \ket{\vec{b}}\bra{\vec{b'}} \sum_{\vec{d}\in \mathcal{D}(\vec{b},\vec{b'})} e^{\mathrm{i}\frac{2\pi}{M}(j-1)\sum_{k=1}^{N} (k-1)(d_k-b_k\oplus d_k \oplus b_k')} \nonumber\\
	\fl &\times \left(\frac{t}{M}\right)^{|\vec{d}|} (1-t)^{|\vec{b}|-|\vec{d}|} \sum_{\vec{f}\in \mathcal{F}(\vec{d})}^{2^N -1} \sum_{\vec{f'}\in \mathcal{F'}(\vec{f},\vec{d},\vec{b},\vec{b'})}^{2^N -1} |\vec{f}|!\,\,(|\vec{d}|-|\vec{f}|)! (\cos\theta)^{2|\vec{f}|-d_1-d_1\oplus b_1 \oplus b_1'} (\sin\theta)^{2(|\vec{d}|-|\vec{f}|)}   \,\,. \label{rho-j-comput8} 
	\end{eqnarray}
	where the sets of binary vectors $\mathcal{D}(\vec{b},\vec{b'})$, $\mathcal{F}(\vec{d})$ and $\mathcal{F'}(\vec{f},\vec{d},\vec{b},\vec{b'})$ are defined as follows:
	\begin{eqnarray}
	\fl&\mathcal{D}(\vec{b},\vec{b'})=\{\vec{d}\in g^{-1}([1,2^N-1]) : \vec{d}\wedge\vec{b}=\vec{d},\,\,\vec{d} \neq \vec{b}\oplus\vec{b'},\,\,(\vec{b}\oplus\vec{d})\wedge \vec{b'}=\vec{b}\oplus\vec{d} \}    \label{set-D}\\
	\fl&\mathcal{F}(\vec{d})=\{\vec{f}\in g^{-1}([0,2^N-1]) : \vec{f}\wedge\vec{d}=\vec{f},\,\,f_1=d_1 \}    \label{set-F}\\
	\fl&\mathcal{F'}(\vec{f},\vec{d},\vec{b},\vec{b'})=\{\vec{f'}\in g^{-1}([0,2^N-1]) : \vec{f'}\wedge(\vec{b}\oplus\vec{d}\oplus\vec{b'})=\vec{f'},\,\,f_1'=b_1 \oplus d_1 \oplus b_1',\,\,|\vec{f'}|=|\vec{f}| \}  \,.  \label{set-F'}
	\end{eqnarray}
	We can now sum over the vectors $\vec{f'}$ since no term depends on them in (\ref{rho-j-comput8}):
	\begin{eqnarray}
	\fl &p_j \rho_{A_1\dots A_N}^j = \nonumber\\
	\fl &\sum_{\stackrel[|\vec{b'}|=|\vec{b}|]{g(\vec{b}),g(\vec{b'})=1}{}}^{2^N-1} q^{N-|\vec{b}|} (1-q)^{|\vec{b}|}  e^{\mathrm{i}[(b_1'-b_1)]\phi} \ket{\vec{b}}\bra{\vec{b'}} \sum_{\vec{d}\in \mathcal{D}(\vec{b},\vec{b'})} e^{\mathrm{i}\frac{2\pi}{M}(j-1)\sum_{k=1}^{N} (k-1)(d_k-b_k\oplus d_k \oplus b_k')} \left(\frac{t}{M}\right)^{|\vec{d}|} (1-t)^{|\vec{b}|-|\vec{d}|} \nonumber\\
	\fl &\times   \sum_{\vec{f}\in \mathcal{F}(\vec{d})}^{2^N -1} {|\vec{b}\oplus\vec{d}\oplus\vec{b'}| - b_1 \oplus d_1 \oplus b_1' \choose |\vec{f}| - b_1 \oplus d_1 \oplus b_1'}^* |\vec{f}|!\,\,(|\vec{d}|-|\vec{f}|)! (\cos\theta)^{2|\vec{f}|-d_1-d_1\oplus b_1 \oplus b_1'} (\sin\theta)^{2(|\vec{d}|-|\vec{f}|)}   \,\,, \label{rho-j-comput9} 
	\end{eqnarray}
	where the asterisk on the binomial coefficient means that it is defined as zero if $|\vec{f}|=0$ and $b_1 \oplus d_1 \oplus b_1'=1$. Finally, since every term just depends on $|\vec{f}|$, we can sum over all the vectors $\vec{f}$ with equal Hamming weight and obtain the final expression for the conditional state of the qubits when detector $D_j$ clicked:
	\begin{eqnarray}
	\fl &p_j \rho_{A_1\dots A_N}^j = \nonumber\\
	\fl &\sum_{\stackrel[|\vec{b'}|=|\vec{b}|]{g(\vec{b}),g(\vec{b'})=1}{}}^{2^N-1} q^{N-|\vec{b}|} (1-q)^{|\vec{b}|}  e^{\mathrm{i}[(b_1'-b_1)]\phi} \ket{\vec{b}}\bra{\vec{b'}} \sum_{\vec{d}\in \mathcal{D}(\vec{b},\vec{b'})} e^{\mathrm{i}\frac{2\pi}{M}(j-1)\sum_{k=1}^{N} (k-1)(d_k-b_k\oplus d_k \oplus b_k')} \left(\frac{t}{M}\right)^{|\vec{d}|} (1-t)^{|\vec{b}|-|\vec{d}|} \nonumber\\
	\fl &\times   \sum_{m=d_1}^{|\vec{d}|} {|\vec{d}| - d_1 \choose m - d_1} {|\vec{b}\oplus\vec{d}\oplus\vec{b'}| - b_1 \oplus d_1 \oplus b_1' \choose m - b_1 \oplus d_1 \oplus b_1'}^* m!\,\,(|\vec{d}|-m)! (\cos\theta)^{2m-d_1-d_1\oplus b_1 \oplus b_1'} (\sin\theta)^{2(|\vec{d}|-m)}   \,\,. \label{rho-j-final} 
	\end{eqnarray}
	
	\subsection{Probability of exactly one click} \label{pj_computation}
	We can now compute the probability $p_j$ of having just one click in detector $D_j$ by simply computing the trace of both sides in (\ref{rho-j-final}):
	\begin{eqnarray}
	\fl p_j = &\sum_{g(\vec{b})=1}^{2^N-1} q^{N-|\vec{b}|} (1-q)^{|\vec{b}|}  \sum_{g(\vec{d})=1\,:\,\vec{d}\wedge\vec{b}=\vec{d}}^{2^N -1} \left(\frac{t}{M}\right)^{|\vec{d}|} (1-t)^{|\vec{b}|-|\vec{d}|}  \nonumber\\
	\fl &\times \sum_{m=d_1}^{|\vec{d}|} \left[{|\vec{d}| - d_1 \choose m - d_1}\right]^2  m!\,\,(|\vec{d}|-m)! (\cos\theta)^{2(m-d_1)} (\sin\theta)^{2(|\vec{d}|-m)}   \,\,.\label{pj-comput1} 
	\end{eqnarray}
	In order to obtain an easier expression to compute, we distinguish the cases: $b_1=0$, $b_1=1$ and the special case $\vec{b}=100\dots 0$:
	\begin{eqnarray}
	\fl &p_j = \sum_{\stackrel[b_1=0]{g(\vec{b})=2}{}}^{2^N-1} q^{N-|\vec{b}|} (1-q)^{|\vec{b}|}  \sum_{\stackrel[d_1=0]{g(\vec{d})=2\,:\,\vec{d}\wedge\vec{b}=\vec{d}}{}}^{2^N -1}  \left(\frac{t}{M}\right)^{|\vec{d}|} (1-t)^{|\vec{b}|-|\vec{d}|} \sum_{m=0}^{|\vec{d}|} \left[{|\vec{d}| \choose m }\right]^2  m!\,(|\vec{d}|-m)! (\cos\theta)^{2m} (\sin\theta)^{2(|\vec{d}|-m)} \nonumber\\
	\fl &+ q^{N-1}(1-q) \left(\frac{t}{M}\right)  \quad\quad \mbox{(this term comes from: $\vec{b}=100\dots 0\,,\,\vec{d}=100\dots 0$ )} \nonumber\\
	\fl &+\sum_{\stackrel[b_1=1]{g(\vec{b})=3}{}}^{2^N-1} q^{N-|\vec{b}|} (1-q)^{|\vec{b}|} \nonumber\\
	\fl &\times\left[ \sum_{\stackrel[d_1=1]{g(\vec{d})=1\,:\,\vec{d}\wedge\vec{b}=\vec{d}}{}}^{2^N -1}  \left(\frac{t}{M}\right)^{|\vec{d}|} (1-t)^{|\vec{b}|-|\vec{d}|} \sum_{m=1}^{|\vec{d}|} \left[{|\vec{d}|-1 \choose m-1}\right]^2  m!\,(|\vec{d}|-m)! (\cos\theta)^{2(m-1)} (\sin\theta)^{2(|\vec{d}|-m)} \right. \nonumber\\
	\fl &\left.+ \sum_{\stackrel[d_1=0]{g(\vec{d})=2\,:\,\vec{d}\wedge\vec{b}=\vec{d}}{}}^{2^N -1}  \left(\frac{t}{M}\right)^{|\vec{d}|} (1-t)^{|\vec{b}|-|\vec{d}|} \sum_{m=0}^{|\vec{d}|} \left[{|\vec{d}| \choose m}\right]^2  m!\,(|\vec{d}|-m)! (\cos\theta)^{2m} (\sin\theta)^{2(|\vec{d}|-m)} \right] \,\,. \label{pj-comput2} 
	\end{eqnarray}
	We can now partially sum over the vectors $\vec{d}$ since the terms in the sums only depend on the Hamming weight of these vectors:
	\begin{eqnarray}
	\fl &p_j = \sum_{\stackrel[b_1=0]{g(\vec{b})=2}{}}^{2^N-1} q^{N-|\vec{b}|} (1-q)^{|\vec{b}|}  \sum_{|\vec{d}|=1}^{|\vec{b}|} {|\vec{b}| \choose |\vec{d}|} \left(\frac{t}{M}\right)^{|\vec{d}|} (1-t)^{|\vec{b}|-|\vec{d}|}  \nonumber\\
	\fl &\times \sum_{m=0}^{|\vec{d}|} \left[{|\vec{d}| \choose m }\right]^2  m!\,(|\vec{d}|-m)! (\cos\theta)^{2m} (\sin\theta)^{2(|\vec{d}|-m)} \,\,+ q^{N-1}(1-q) \left(\frac{t}{M}\right)   \nonumber\\
	\fl &+\sum_{\stackrel[b_1=1]{g(\vec{b})=3}{}}^{2^N-1} q^{N-|\vec{b}|} (1-q)^{|\vec{b}|} \nonumber\\
	\fl &\times\left[ \sum_{|\vec{d}|=1}^{|\vec{b}|}  {|\vec{b}|-1 \choose |\vec{d}| -1}\left(\frac{t}{M}\right)^{|\vec{d}|} (1-t)^{|\vec{b}|-|\vec{d}|} \sum_{m=1}^{|\vec{d}|} \left[{|\vec{d}|-1 \choose m-1}\right]^2  m!\,(|\vec{d}|-m)! (\cos\theta)^{2(m-1)} (\sin\theta)^{2(|\vec{d}|-m)} \right. \nonumber\\
	\fl &\left.+ \sum_{|\vec{d}|=1}^{|\vec{b}|-1}  {|\vec{b}|-1 \choose |\vec{d}|} \left(\frac{t}{M}\right)^{|\vec{d}|} (1-t)^{|\vec{b}|-|\vec{d}|} \sum_{m=0}^{|\vec{d}|} \left[{|\vec{d}| \choose m}\right]^2  m!\,(|\vec{d}|-m)! (\cos\theta)^{2m} (\sin\theta)^{2(|\vec{d}|-m)} \right] \,\,. \label{pj-comput3} 
	\end{eqnarray}
	By employing the following identity:
	\begin{eqnarray}
	\fl &\sum_{m=0}^{|\vec{d}|} \left[{|\vec{d}| \choose m}\right]^2  m!\,(|\vec{d}|-m)! (\cos\theta)^{2m} (\sin\theta)^{2(|\vec{d}|-m)} =  |\vec{d}|! \,\sum_{m=0}^{|\vec{d}|} {|\vec{d}| \choose m} (\cos\theta)^{2m} (\sin\theta)^{2(|\vec{d}|-m)} \nonumber\\
	\fl &= |\vec{d}|! \, \left(\sin^2 \theta + \cos^2 \theta \right)^{|\vec{d}|} = |\vec{d}|! \label{identity} \,\,,
	\end{eqnarray}
	both in the second and last row of (\ref{pj-comput3}), we get the following expression:
	\begin{eqnarray}
	\fl &p_j = \sum_{\stackrel[b_1=0]{g(\vec{b})=2}{}}^{2^N-1} q^{N-|\vec{b}|} (1-q)^{|\vec{b}|}  \sum_{l=1}^{|\vec{b}|} {|\vec{b}| \choose l} \left(\frac{t}{M}\right)^l (1-t)^{|\vec{b}|-l} l! \, + \, q^{N-1}(1-q) \left(\frac{t}{M}\right)   \nonumber\\
	\fl &+\sum_{\stackrel[b_1=1]{g(\vec{b})=3}{}}^{2^N-1} q^{N-|\vec{b}|} (1-q)^{|\vec{b}|} \nonumber\\
	\fl &\times\left[ \sum_{l=0}^{|\vec{b}|-1}  {|\vec{b}|-1 \choose l}\left(\frac{t}{M}\right)^{l+1} (1-t)^{|\vec{b}|-l-1} \sum_{m=0}^{l} \left[{l \choose m}\right]^2  (m+1)!\,(l-m)! (\cos\theta)^{2m} (\sin\theta)^{2(l-m)} \right. \nonumber\\
	\fl &\left.+ \sum_{l=1}^{|\vec{b}|-1}  {|\vec{b}|-1 \choose l} \left(\frac{t}{M}\right)^{l} (1-t)^{|\vec{b}|-l} l! \right] \,\,. \label{pj-comput4} 
	\end{eqnarray}
	The remaining sum over $m$ can be similarly simplified as follows:
	\begin{eqnarray}
	\sum_{m=0}^{l} \left[{l \choose m}\right]^2  (m+1)!\,(l-m)! (\cos\theta)^{2m} (\sin\theta)^{2(l-m)} = l! \, \left(1+l \cos^2 \theta \right) \,\,. \label{identity2}
	\end{eqnarray}
	By substituting (\ref{identity2}) into (\ref{pj-comput4}) and by partially summing over vectors $\vec{b}$ we obtain the final expression for the probability that only detector $D_j$ clicks:
	\begin{eqnarray}
	\fl p_j = &N q^{N-1}(1-q) \left(\frac{t}{M}\right) \,+\, \sum_{r=2}^{N-1} {N-1 \choose r} q^{N-r} (1-q)^r \sum_{l=1}^{r} {r \choose l} \left(\frac{t}{M}\right)^l (1-t)^{r-l} l!    \nonumber\\
	\fl &+\sum_{r=2}^{N} {N-1 \choose r-1} q^{N-r} (1-q)^r \left[ \sum_{l=1}^{r}  {r-1 \choose l-1}\left(\frac{t}{M}\right)^{l} (1-t)^{r-l} (l-1)!(\sin^2 \theta + l \cos^2 \theta) \right. \nonumber\\
	\fl &\left.+ \sum_{l=1}^{r-1}  {r-1 \choose l} \left(\frac{t}{M}\right)^{l} (1-t)^{r-l} l! \right] \label{pj} \,\,.
	\end{eqnarray}
	We observe that the probability $p_j$ that only detector $D_j$ clicks is independent of the particular detector because of the symmetric action of the multiport beam splitter.
	
	\subsection{QBER}
	Starting from the conditional state (\ref{rho-j-final}), we can compute the QBER between $\mathrm{Alice}_1$ and $\mathrm{Alice}_k$'s outcomes when they measure their qubit in the eigenbasis of the operator $O_{XY}(\varphi_1)$ and $O_{XY}(\varphi_k)$, respectively, with $O_{XY}(\varphi)=\cos(\varphi) X + \sin(\varphi) Y$ ($X$ and $Y$ are the Pauli operators). The operator $O_{XY}(\varphi)$ has eigenvalues $\lambda=\pm 1$ and correspondent eigenvectors: $\ket{\lambda}_{\varphi}=\frac{1}{\sqrt{2}} (\ket{0}+\lambda e^{\mathrm{i} \varphi} \ket{1})$.\\
	To start with, we compute the following quantity:
	\begin{eqnarray}
	\fl \bra{\vec{b'}} P^{A_1}_{\ket{+1}_{\varphi_1}} \otimes P^{A_k}_{\ket{-1}_{\varphi_k}} \ket{\vec{b}} = \frac{1}{4} \left(e^{{b'}_1\, \mathrm{i}\varphi_1}\right) \left((-1)^{{b'}_k}e^{{b'}_k\, \mathrm{i}\varphi_k}\right) \left(e^{-{b}_1\, \mathrm{i}\varphi_1}\right) \left((-1)^{{b}_k}e^{-{b}_k\, \mathrm{i}\varphi_k}\right) \,\prod_{l\neq 1,k} \delta_{b_l,{b'}_l} \,\, \label{QBER-comput1}
	\end{eqnarray}
	and we insert it into the probability that $\mathrm{Alice}_1$ measured the outcome +1 and $\mathrm{Alice}_k$ measured the outcome -1, when $D_j$ clicked:
	\begin{eqnarray}
	\fl &\Tr \left[P^{A_1}_{\ket{+1}_{\varphi_1}} \otimes P^{A_k}_{\ket{-1}_{\varphi_k}} \rho^j_{A_1,\dots,A_N}\right] = \frac{1}{p_j} \sum_{g(\vec{b})=1}^{2^N-1} \sum_{\vec{b'}\in \mathcal{Q}(\vec{b})} \frac{(-1)^{b_k +{b'}_k}}{4} e^{\mathrm{i}({b'}_1 -b_1)(\varphi_1+\phi)} e^{({b'}_k -b_k)\mathrm{i}\varphi_k} q^{N-|\vec{b}|} (1-q)^{|\vec{b}|} \nonumber\\
	\fl &\times \sum_{\vec{d}\in \mathcal{D}(\vec{b},\vec{b'})} \left(\frac{t}{M}\right)^{|\vec{d}|}(1-t)^{|\vec{b}|-|\vec{d}|} e^{\mathrm{i}\frac{2\pi}{M}(j-1)(k-1)(d_k-b_k\oplus {b'}_k\oplus d_k)} \nonumber\\
	\fl &\times \sum_{m=d_1}^{|\vec{d}|} {|\vec{d}| - d_1 \choose m - d_1} {|\vec{b}\oplus\vec{d}\oplus\vec{b'}| - b_1 \oplus d_1 \oplus b_1' \choose m - b_1 \oplus d_1 \oplus b_1'}^* m!\,\,(|\vec{d}|-m)! (\cos\theta)^{2m-d_1-d_1\oplus b_1 \oplus b_1'} (\sin\theta)^{2(|\vec{d}|-m)}  \label{QBER-comput2}
	\end{eqnarray}
	where $\mathcal{Q}(\vec{b})$ is a set of at most 2 binary vectors, defined as: $\mathcal{Q}(\vec{b})=\{\vec{b},\,\overline{b_1} b_2 \dots \overline{b_k} \dots b_N \mbox{(iff $b_1\oplus b_k=1$)}\}$\footnote{The straight line over a bit indicates its negation.}. This means that the sum over $\vec{b'}$ is reduced to just one term, namely $\vec{b'}=\vec{b}$, plus the possibility of a second term in which $\vec{b'}$ differs from $\vec{b}$ in position 1 and $k$, as long as the bits of vector $\vec{b}$ differ from each other in those positions. Now we compute the sum over $\vec{b'}$ as follows:
	\begin{eqnarray}
	\fl &\Tr \left[P^{A_1}_{\ket{+1}_{\varphi_1}} \otimes P^{A_k}_{\ket{-1}_{\varphi_k}} \rho^j_{A_1,\dots,A_N}\right] =\frac{1}{p_j} \Biggg\lbrace \sum_{g(\vec{b})=1}^{2^N-1} \frac{q^{N-|\vec{b}|} (1-q)^{|\vec{b}|}}{4} \sum_{g(\vec{d})=1\,:\,\vec{d}\wedge\vec{b}=\vec{d}}^{2^N -1} \left(\frac{t}{M}\right)^{|\vec{d}|} (1-t)^{|\vec{b}|-|\vec{d}|}  \nonumber\\
	\fl &\times \sum_{m=d_1}^{|\vec{d}|} \left[{|\vec{d}| - d_1 \choose m - d_1}\right]^2  m!\,\,(|\vec{d}|-m)! (\cos\theta)^{2(m-d_1)} (\sin\theta)^{2(|\vec{d}|-m)} \nonumber\\
	\fl &- \sum_{g(\vec{b})=1\,:\,b_1\oplus b_k=1}^{2^N-1} \frac{q^{N-|\vec{b}|} (1-q)^{|\vec{b}|}}{4} e^{\mathrm{i} (-1)^{b_1} (\varphi_1 +\phi) + \mathrm{i}(-1)^{b_k} \varphi_k +\mathrm{i}\frac{2\pi}{M}(j-1)(k-1)(-1)^{b_k \oplus 1}} \nonumber\\
	\fl &\times\sum_{\vec{d}\in \mathcal{D}(\vec{b},\overline{b_1} b_2 \dots \overline{b_k} \dots b_N)} \left(\frac{t}{M}\right)^{|\vec{d}|}(1-t)^{|\vec{b}|-|\vec{d}|}  
	\sum_{m=b_1}^{|\vec{d}|} {|\vec{d}| - b_1 \choose m - b_1}{|\vec{d}| - \overline{b_1} \choose m - \overline{b_1}}^*  m!\,(|\vec{d}|-m)! (\cos\theta)^{2m-1} (\sin\theta)^{2(|\vec{d}|-m)}\Biggg\rbrace,\nonumber\\
	\fl &  \label{QBER-comput3}
	\end{eqnarray}
	where the set $\mathcal{D}(\vec{b},\overline{b_1} b_2 \dots \overline{b_k} \dots b_N)$ simplifies to: $\mathcal{D}(\vec{b},\overline{b_1} b_2 \dots \overline{b_k} \dots b_N)=\{\vec{d}\in g^{-1}([1,2^N-1]) : \vec{d}\wedge\vec{b}=\vec{d},\,\,d_1=b_1,\,\,d_k=b_k\}$. We notice that the first addend in (\ref{QBER-comput3})
	is proportional to $p_j$ (\ref{pj-comput1}):
	\begin{eqnarray}
	\fl &\Tr \left[P^{A_1}_{\ket{+1}_{\varphi_1}} \otimes P^{A_k}_{\ket{-1}_{\varphi_k}} \rho^j_{A_1,\dots,A_N}\right] = \nonumber\\
	\fl &=\frac{1}{4} - \frac{1}{4 p_j} \sum_{g(\vec{b})=1\,:\,b_1\oplus b_k=1}^{2^N-1} q^{N-|\vec{b}|} (1-q)^{|\vec{b}|} e^{\mathrm{i} (-1)^{b_1} (\varphi_1 +\phi) + \mathrm{i}(-1)^{b_k} \varphi_k +\mathrm{i}\frac{2\pi}{M}(j-1)(k-1)(-1)^{b_k \oplus 1}} \nonumber\\
	\fl &\sum_{\vec{d}\in \mathcal{D}(\vec{b},\overline{b_1} b_2 \dots \overline{b_k} \dots b_N)} \left(\frac{t}{M}\right)^{|\vec{d}|}(1-t)^{|\vec{b}|-|\vec{d}|}  \sum_{m=b_1}^{|\vec{d}|} {|\vec{d}| - b_1 \choose m - b_1}{|\vec{d}| - \overline{b_1} \choose m - \overline{b_1}}^*  m!\,(|\vec{d}|-m)! (\cos\theta)^{2m-1} (\sin\theta)^{2(|\vec{d}|-m)}. \nonumber\\
	\fl & \label{QBER-comput4}
	\end{eqnarray}
	Finally, we split the sums over $\vec{b}$ in the two sub-cases: $b_1=1,b_k=0$ and $b_1=0,b_k=1$ and we notice that the two contributions differ only in the exponential term. By summing the two contributions, the exponential factor produces a cosine function and one gets:
	\begin{eqnarray}
	\fl &\Tr \left[P^{A_1}_{\ket{+1}_{\varphi_1}} \otimes P^{A_k}_{\ket{-1}_{\varphi_k}} \rho^j_{A_1,\dots,A_N}\right] = \frac{1}{4} - \frac{2\cos\left[(\varphi_1+\phi-\varphi_k) + \frac{2\pi }{M}(j-1)(k-1)\right]}{4 p_j} \nonumber\\
	\fl &\times \sum_{|\vec{b}|=1}^{N-1} {N-2 \choose |\vec{b}|-1} q^{N-|\vec{b}|}(1-q)^{|\vec{b}|} \sum_{|\vec{d}|=1}^{|\vec{b}|} {|\vec{b}|-1 \choose |\vec{d}|-1} \left(\frac{t}{M}\right)^{|\vec{d}|}(1-t)^{|\vec{b}|-|\vec{d}|} \nonumber\\
	\fl &\times \sum_{m=1}^{|\vec{d}|} {|\vec{d}|  \choose m }{|\vec{d}| - 1 \choose m -1}  m!\,(|\vec{d}|-m)! (\cos\theta)^{2m-1} (\sin\theta)^{2(|\vec{d}|-m)} \nonumber\\
	\fl &= \frac{1}{4} - \frac{\cos\left[(\varphi_1+\phi-\varphi_k) + \frac{2\pi }{M}(j-1)(k-1)\right]}{2 p_j} \nonumber\\
	\fl &\times \sum_{r=0}^{N-2} {N-2 \choose r} q^{N-r-1}(1-q)^{r+1} \sum_{l=0}^{r} {r \choose l} \left(\frac{t}{M}\right)^{l+1}(1-t)^{r-l} (l+1)!  \sum_{m=0}^{l} {l  \choose m } (\cos\theta)^{2m+1} (\sin\theta)^{2(l-m)} \nonumber\\
	\fl &= \frac{1}{4} - \frac{1}{2p_j}\cos\left[(\varphi_1+\phi-\varphi_k) + \frac{2\pi }{M}(j-1)(k-1)\right] \cos\theta \nonumber\\
	\fl &\times \sum_{r=0}^{N-2} {N-2 \choose r} q^{N-r-1}(1-q)^{r+1} \sum_{l=0}^{r} {r \choose l} \left(\frac{t}{M}\right)^{l+1}(1-t)^{r-l} (l+1)! \,\,. \label{QBER-comput5}
	\end{eqnarray}
	In a similar fashion, one computes the probability of $A_1$ measuring the outcome -1 and $A_k$ measuring the outcome +1 and obtains an identical expression to (\ref{QBER-comput5}). In conclusion, the QBER conditioned on $D_j$ clicking is given by twice the probability given in (\ref{QBER-comput5}).\\
	By fixing the angles $\varphi_1$ and $\varphi_k$ as mentioned in the protocol's description: $\varphi_1=0$ and $\varphi_k=\arg(U_{kj})=\frac{2\pi}{M} (k-1)(j-1)$ we minimize the QBER and thus increase the secret key rate. This requires $\mathrm{Alice}_k$ to adjust her measurement depending on which detector clicked, implying that such measurement does not commute with the operations performed by node $C$. On the other hand, the QBER is now minimal and reads the same regardless of which couple $(A_1,A_k)$ one considers or which detector $D_j$ clicks:
	\begin{eqnarray}
	\fl Q_{A_1 A_k}= \frac{1}{2} - \frac{1}{p_j}\cos\phi \cos\theta \sum_{r=0}^{N-2} {N-2 \choose r} q^{N-r-1}(1-q)^{r+1}\sum_{l=0}^{r} {r \choose l} (l+1)! \left(\frac{t}{M}\right)^{l+1} (1-t)^{r-l} \label{QBER}\,\,,
	\end{eqnarray}
	where $p_j$ is given in (\ref{pj}).
	
	\subsection{Phase-error rate}
	Finally we compute the phase-error rate, defined as the probability that the product of the $Z$-measurement results of all the parties equals +1 (i.e. the qubit of an even number of parties collapsed in state $\ket{1}$, which corresponds to the outcome $Z=-1$):
	\begin{eqnarray}
	 Q_Z=  \mathrm{Pr}\left[\prod_{k=1}^{N} Z_{A_k}=1\right] = \Tr \Bigg[\Bigg(\sum_{\stackrel[|\vec{f}|\,\mathrm{even}]{g(\vec{f})=3}{}}^{2^N-1} P_{\ket{\vec{f}}}\Bigg)  \rho^j_{A_1,\dots,A_N} \Bigg] \label{phase-error-rate-comput1}
	\end{eqnarray}
	where the quantum state $\rho^j_{A_1,\dots,A_N}$ conditioned on detector $D_j$ clicking is given in (\ref{rho-j-final}) and the case $g(\vec{f})=0$ is excluded since $\ket{\vec{0}}$ does not appear in (\ref{rho-j-final}). By following analogous steps to those presented in \ref{pj_computation} we obtain the following expression for the phase-error rate:
	\begin{eqnarray}
	\fl Q_Z = &\frac{1}{p_j}\sum_{r=1}^{\left\lfloor\frac{N-1}{2}\right\rfloor} {N-1 \choose 2r} q^{N-2r} (1-q)^{2r} \sum_{l=1}^{2r} {2r \choose l} \left(\frac{t}{M}\right)^l (1-t)^{2r-l} l!    \nonumber\\
	\fl &+\frac{1}{p_j}\sum_{r=1}^{\left\lfloor\frac{N}{2}\right\rfloor} {N-1 \choose 2r-1} q^{N-2r} (1-q)^{2r} \left[ \sum_{l=1}^{2r-1}  {2r-1 \choose l}\left(\frac{t}{M}\right)^{l} (1-t)^{2r-l} l! \right. \nonumber\\
	\fl &\left.+ \sum_{l=1}^{2r}  {2r-1 \choose l-1} \left(\frac{t}{M}\right)^{l} (1-t)^{2r-l} (l-1)! (\sin^2 \theta + l \cos^2 \theta) \right] \label{QZ} \,\,.
	\end{eqnarray}
	
	\subsection{Dark counts}
	So far we computed the quantities $p_j$, $Q_{A_1 A_k}$ and $Q_Z$ assuming that every click in the detectors is due to the arrival of one or more photons. By naming $\Omega_{\mathrm{ph}}$ the event in which one or more photons arrive at detector $D_j$ and no other photon arrives at any other detector, we can formally express the computed quantities as:
	\begin{eqnarray}
		p_j &= \mathrm{Pr}\left(\Omega_{\mathrm{ph}}\right) \label{pj-prob} \\
		Q_{A_1 A_k} &= \mathrm{Pr}\left(A_1 \neq A_k | \Omega_{\mathrm{ph}}\right) \label{QBER-prob} \\
		Q_Z &= \mathrm{Pr}\left(\textstyle\prod_{k=1}^{N} Z_{A_k}=1 | \Omega_{\mathrm{ph}}\right) \,\,. \label{QZ-prob} 
	\end{eqnarray}
	For the setup presented in \autoref{CKA-protocol} and the channel model described at the beginning of this Section, the explicit expressions of (\ref{pj-prob}), (\ref{QBER-prob}) and (\ref{QZ-prob}) are given in (\ref{pj}), (\ref{QBER}) and (\ref{QZ}), respectively.\\
	We now assume that every detector is characterized by a probability $p_d$ of clicking conditioned on no photon arriving. We also define $\Omega_{\mathrm{click}}$ to be the event in which only detector $D_j$ clicks and $\Omega_{\mathrm{no\,ph}}$ to be the event in which no photon arrives at any detector. Then, the error rates $Q^{\mathrm{dc}}_{A_1 A_k}$ and $Q_Z^{\mathrm{dc}}$ and the probability $p^{\mathrm{dc}}_j$ that enter the key rate formula and that model the correspondent observed quantities read as follows:
	\begin{eqnarray}
	\fl p^{\mathrm{dc}}_j &= \mathrm{Pr}\left(\Omega_{\mathrm{click}}\right)= p_j (1-p_d)^{M-1} +p_d (1-p_d)^{M-1} \mathrm{Pr}(\Omega_{\mathrm{no\,ph}}) \label{pj-dc} \\
	\fl Q^{\mathrm{dc}}_{A_1 A_k} &= \mathrm{Pr}\left(A_1 \neq A_k | \Omega_{\mathrm{click}}\right) \nonumber\\
	\fl &=\frac{1}{p_j^{\mathrm{dc}}}\left[\mathrm{Pr}\left(A_1 \neq A_k | \Omega_{\mathrm{no\,ph}}\right)\mathrm{Pr}\left(\Omega_{\mathrm{no\,ph}}\right)p_d (1-p_d)^{M-1}+ Q_{A_1 A_k}\, p_j (1-p_d)^{M-1} \right] \label{QBER-dc} \\
	\fl Q^{\mathrm{dc}}_Z &= \mathrm{Pr}\left(\textstyle\prod_{k=1}^{N} Z_{A_k}=1 | \Omega_{\mathrm{click}}\right)  \nonumber\\
	\fl &=\frac{1}{p_j^{\mathrm{dc}}}\left[\mathrm{Pr}\left(\textstyle\prod_{k=1}^{N} Z_{A_k}=1|\Omega_{\mathrm{no\,ph}}\right)\mathrm{Pr}\left(\Omega_{\mathrm{no\,ph}}\right) p_d (1-p_d)^{M-1} + Q_Z \, p_j (1-p_d)^{M-1} \right]  \,\,, \label{QZ-dc} 
	\end{eqnarray}
	where $p_j$, $Q_{A_1 A_k}$ and $Q_Z$ are defined in (\ref{pj-prob}), (\ref{QBER-prob}) and (\ref{QZ-prob}), respectively, while the probabilities related to the arrival of no photon read:
	\begin{eqnarray}
		&\mathrm{Pr}\left(\Omega_{\mathrm{no\,ph}}\right) = \left(q + (1-q)(1-t)\right)^N  \label{no_ph-prob} \\
		&\mathrm{Pr}\left(A_1 \neq A_k | \Omega_{\mathrm{no\,ph}}\right) = \frac{1}{2} \label{QBER-no_ph} \\
		&\mathrm{Pr}\left(\textstyle\prod_{k=1}^{N} Z_{A_k}=1|\Omega_{\mathrm{no\,ph}}\right) = \frac{1}{\mathrm{Pr}\left(\Omega_{\mathrm{no\,ph}}\right)}\sum_{l=0}^{\left\lfloor\frac{N}{2}\right\rfloor} {N \choose 2l} q^{N-2l}(1-q)^{2l}(1-t)^{2l}  \label{QZ-no_ph} \,\,.
	\end{eqnarray}
	The probabilities (\ref{no_ph-prob}), (\ref{QBER-no_ph}) and (\ref{QZ-no_ph}) are obtained by following similar steps to those presented in this Section and that led to the final expressions for $p_j$, $Q_{A_1 A_k}$ and $Q_Z$, respectively. The starting point in this case is the conditional state of the qubits when no photon arrived at any detector:
	\begin{eqnarray}
		\mathrm{Pr}\left(\Omega_{\mathrm{no\,ph}}\right) \rho_{A_1\dots A_N}^{\mathrm{no\,ph}} = \Tr_{\stackrel[l_1,\dots,l_N]{\sigma_1,\dots,\sigma_M}{}}\left[\otimes_{i=1}^M P_{\ket{0}_{\sigma_i}} \ket{\Phi_4}\bra{\Phi_4}\otimes_{i=1}^M P_{\ket{0}_{\sigma_i}} \right] \label{rho-no_ph}
	\end{eqnarray}
	
	\section{Optimized conference key agreement} \label{optimizedCKA}
	\begin{figure}[!htb]
		\centering
		\begin{subfigure}[t]{.5\textwidth}
			\centering
			\includegraphics[width=1\linewidth,keepaspectratio]{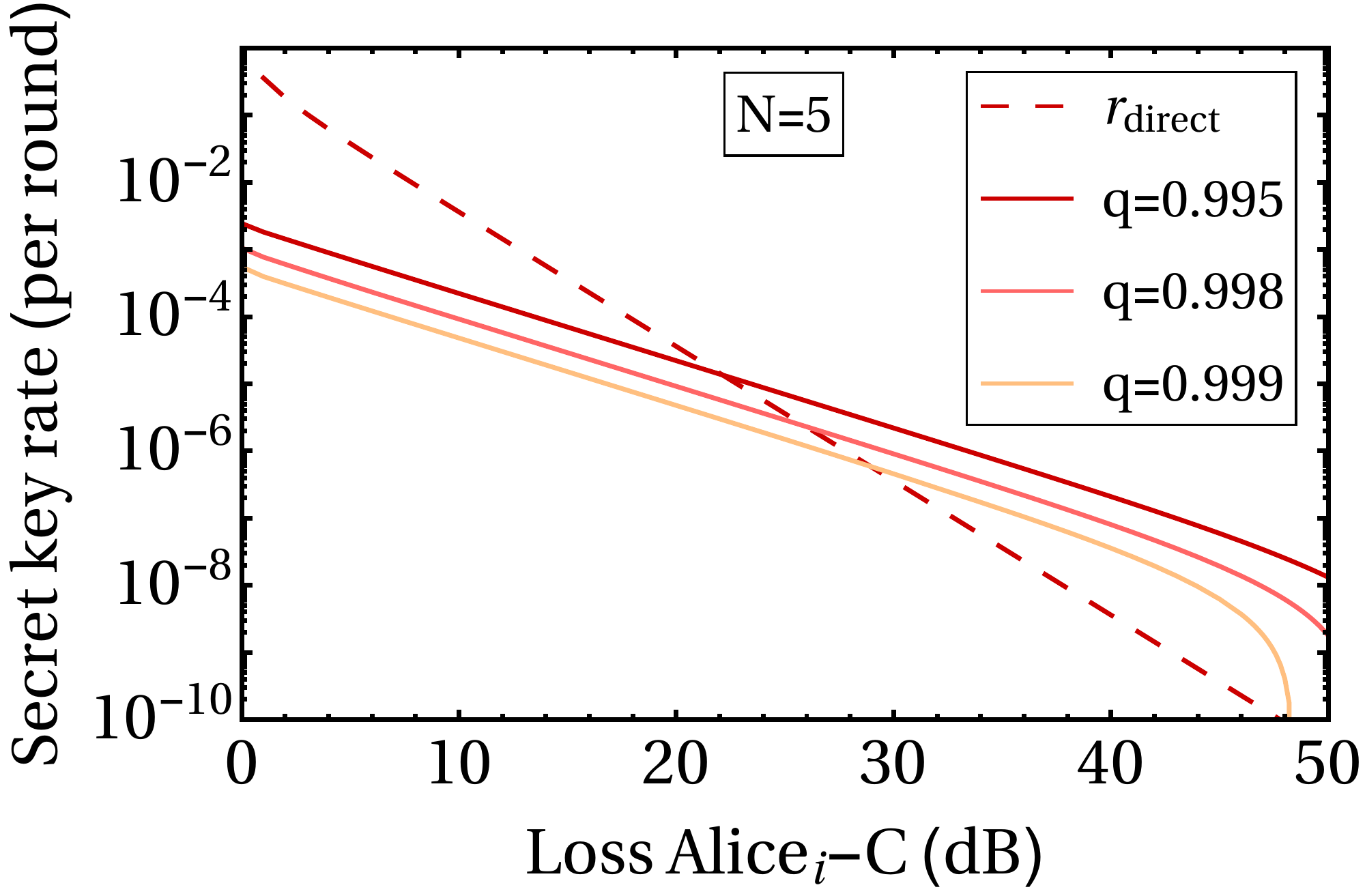}
			\caption{The optimized conference key rate (Eq.~\ref{opt-key-rate}, solid lines) as a function of the loss in the channel linking one party to the central node, for different fixed values of $q$: $q=0.995,0.998$ and $0.999$ (top to bottom). We also plot the direct transmission bound (Eq.~\ref{PLOB}, dashed line) for five parties. We observe that the optimized key rate outperforms the standard CKA especially at high losses (compare with the case $N=5$ in \autoref{asymptotic_plot}), since having a lower number of parties taking part to the CKA all at once reduces the negative effect of dark counts.}
			\label{CKA-fixed_q}
		\end{subfigure}%
		\begin{subfigure}[t]{.5\textwidth}
			\centering
			\includegraphics[width=1\linewidth,keepaspectratio]{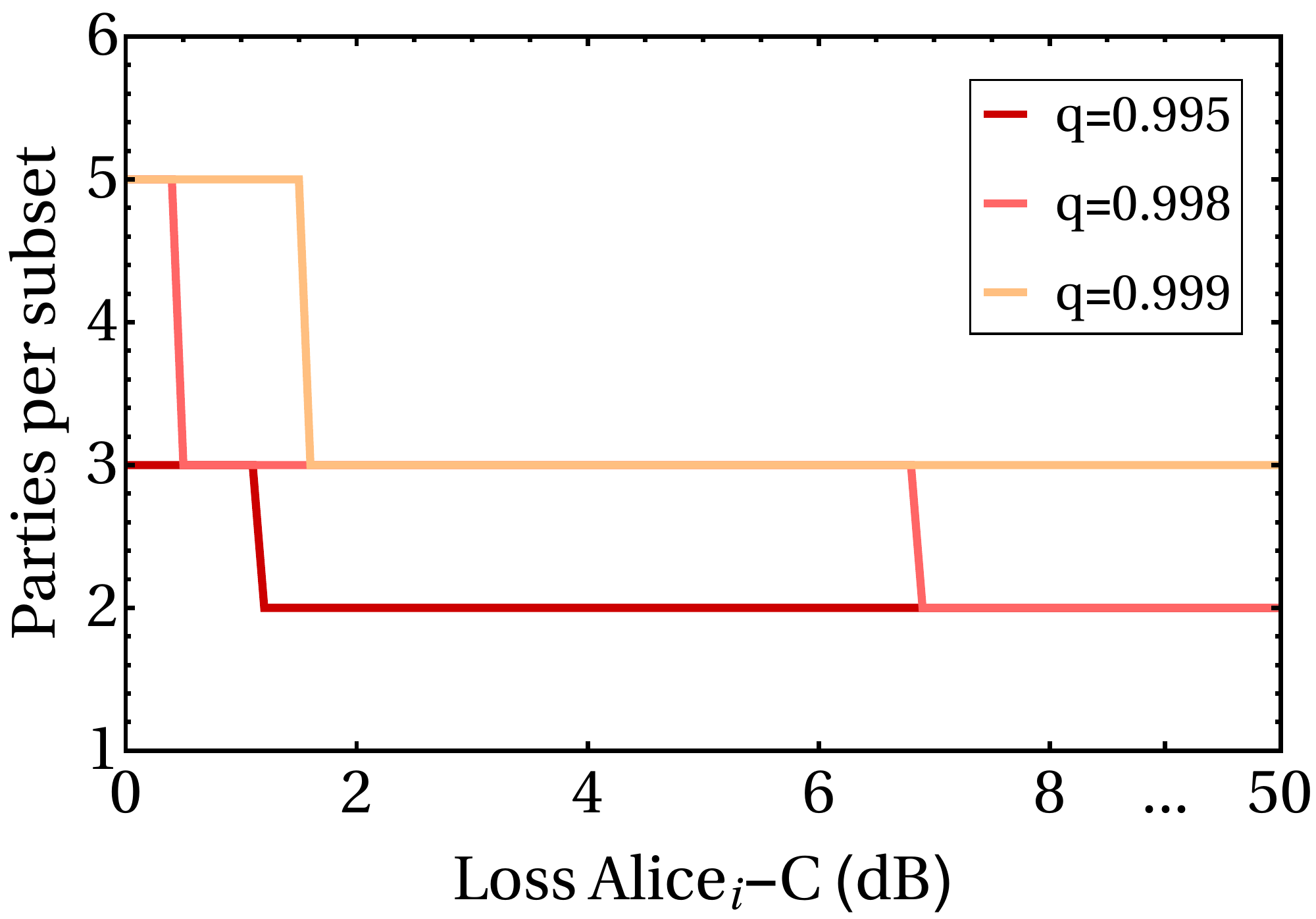}
			\caption{The optimal number of parties belonging to the subsets in which the total number of users ($N=5$) have been subdivided, as a function of the loss in one quantum channel, for different fixed values of $q$: $q=0.995,0.998$ and $0.999$ (bottom to top). We observe that performing a truly multipartite scheme could be optimal especially at low losses, i.e. when the parties' shared state is well approximated by a multipartite $W$ state.}
			\label{optimal-number-parties}
		\end{subfigure}
	\caption{We optimize the conference key rate achieved by $N=5$ parties over the cardinality of the subsets of parties performing the CKA all at once. The different keys established within each subgroup (which can be composed of either two, three or five parties each) are then used to encode the final conference key.}
	\label{optimized-subgroups-rate}
	\end{figure}
	Although we have shown in \autoref{simulations} that a truly multipartite QKD scheme can outperform the iterative use of any bipartite QKD scheme in the direct-transmission scenario (i.e. the central node is removed), this does not necessarily hold when one has at hand the CKA experimental setup and uses it to perform bipartite QKD protocols. In other words, it might be possible to outperform the multipartite CKA by iteratively executing its bipartite version (fix $N=2$ in Eq.~\ref{key-rate}) between one selected party and all the other $N-1$ users, and then using the established secret keys to encode the final conference key via one-time pad encryption. In this case the asymptotic conference key rate would be $r(2)/(N-1)$ according to the reasoning given at the beginning of \autoref{simulations}, where $r(N)$ is given in (\ref{key-rate}). More generally, it might be advantageous to group the $N$ parties in subsets of equal cardinality, let them perform the CKA within the subset, and then use the secret keys established in each subset to encode the conference key. Since one selected party must belong to every subset in order to distribute the final conference key to the others in a secure way, there are $(N-1)/d$ subsets of $d+1$ users each. In this case the asymptotic conference key rate would read: $d\cdot r(d+1)/(N-1)$. In order to investigate which of these configurations yields the highest asymptotic conference key rate, we optimize the rate with respect to the possible subdivisions of the $N$ parties in groups of equal cardinality (i.e. we maximize it with respect to all the divisors $d$ of $N-1$):
	\begin{equation}
	r_{\mathrm{opt}}(N)=\max_{d|N-1} \frac{d}{N-1} r(d+1) \,\,,  \label{opt-key-rate}
	\end{equation}
	which includes the cases where the parties are iteratively performing bipartite protocols ($d=1$) and where the $N$ parties are performing the CKA all at once like in \autoref{asymptotic_plot} ($d=N-1$).\\
	In \autoref{CKA-fixed_q} we plot the optimized conference key rate for $N=5$ parties (Eq.~\ref{opt-key-rate}, solid lines) as a function of the loss in each quantum channel, for different fixed values of the parameter $q$ and a fixed number of input (output) ports of the beam splitter: $M=5$. We also plot the direct transmission bound (\ref{PLOB}) for the same number of parties. The correspondent optimal number of parties within each subset depends on the loss and on the value of $q$, and it is given in \autoref{optimal-number-parties}.\\
	From \autoref{CKA-fixed_q} we observe that the resulting key rate, although not being optimized over the parameter $q$, is similar to the $N=5$ key rate in \autoref{asymptotic_plot} for most losses, since we fixed $q$ to values close to the optimal ones. Furthermore, it performs better than the standard CKA with five parties in the high-loss region. Indeed, as already explained in \autoref{asymptotic_plot}, the effect of dark counts becomes greater when more parties are performing the CKA at the same time. Thus, allowing for a lower number of parties within each subset increases the maximum tolerated loss. \\
	In \autoref{optimal-number-parties} we observe that at low losses it is optimal for the five parties to perform a truly multipartite scheme rather than iteratively performing bipartite protocols. The reason is that in the ideal scenario of extremely low losses ($t\longrightarrow 1$) and $q$ close to 1, there is only one party successfully sending one photon to the central node to be detected. In this case the post-selected state shared by the parties is the $W$-class state used for establishing the secret key. Of course, there are more chances that this event is going to happen when more parties are involved, thus a multipartite scheme is advantageous with respect to an iteration of bipartite schemes. One can see this also analytically, by showing that the asymptotic rate (\ref{key-rate}) of the CKA performed by $N$ parties all at once can be approximated as follows (when the above assumptions hold):
	\begin{equation}
	r(N)  \simeq N q^{N-1} (1-q) t \left[1-h\left(\frac{1}{2} - \frac{1}{N}\right)\right]\,\,,  \label{approx-key-rate}
	\end{equation}
	while the rate achieved by subdividing the task in $N-1$ bipartite schemes is:
	\begin{equation}
	r_{\mathrm{bipartite}}(N) \simeq \frac{2 q (1-q) t}{N-1}  \label{bipartite} \,\,.
	\end{equation}
	By numerically comparing (\ref{approx-key-rate}) with (\ref{bipartite}) for sufficiently high values of $q$, one notices that the former results in a higher key rate. When the value of $q$ decreases, the probability that two or more parties send their photon to the central node increases, reducing the key rate. Being such events more likely when more parties are involved, the iterative execution of bipartite schemes is favored. 
	Similarly, increasing the loss transforms the same events --which are more likely with more parties-- from neglected events (if they cause double clicks) to harmful events (when some photons get lost in the transmission), thus favoring the iteration of schemes with a low number of parties.


\section*{References}
\begin{thebibliography}{99}
	\small
	\bibitem{quantum-communication-review} N. Gisin and R. Thew. \textit{Nat. Photon.} \textbf{1}, 165–171 (2007).
	\bibitem{Kimble} H. J. Kimble. \textit{Nature} \textbf{453}, 1023 (2008).
	\bibitem{BB84} C. H. Bennett and G. Brassard. \textit{Proc. IEEE Int. Conf. on Computers, Systems and Signal Processing}, pp 175–9 (1984).
	\bibitem{E91} A. K. Ekert. \textit{Phys. Rev. Lett.} \textbf{67}, 661 (1991).
	\bibitem{Scarani-review} V. Scarani, H. Pasquinucci, N. J. Cerf, M. Dušek, N. Lütkenhaus, and M. Peev. \textit{Rev. Mod. Phys.} \textbf{81}, 1301 (2009).
	\bibitem{Curty-review} H.-K. Lo, M. Curty, and K. Tamaki. \textit{Nat. Photon.} \textbf{8}, 595–604 (2014).
	\bibitem{Diamanti-review} E. Diamanti, H.-K. Lo, B. Qi and Z. Yuan. \textit{npj Quantum Inf.} \textbf{2}, 16025 (2016).
	\bibitem{Pirandola-review} S. Pirandola \textit{et al.} \textit{preprint arXiv}:1906.01645.
	\bibitem{Curty-MDIQKD} H.-K. Lo, M. Curty, and B. Qi. \textit{Phys. Rev. Lett.} \textbf{108}, 130503 (2012).
	\bibitem{Abruzzo-MDIQKD} S. Abruzzo, H. Kampermann, and D. Bruß. \textit{Phys. Rev. A} \textbf{89}, 012301 (2014).
	\bibitem{mem-assisited-MDIQKD} C. Panayi, M. Razavi, X. Ma, and N. Lütkenhaus. \textit{New J. Phys.} \textbf{16}, 043005 (2014).
	\bibitem{Azuma-intercityQKD} K. Azuma, K. Tamaki, and W. J. Munro. \textit{Nat. Comm.} \textbf{6}, 10171 (2015).
	\bibitem{DIQKD1} U. Vazirani and T. Vidick. \textit{Phys. Rev. Lett.} \textbf{113}, 140501 (2014).
	\bibitem{DIQKD2} R. A. Friedman, F. Dupuis, O. Fawzi, R. Renner, and T. Vidick.  \textit{Nat. Commun.} \textbf{9}, 459 (2018).
	\bibitem{404km} H.-L. Yin et al.. \textit{Phys. Rev. Lett.} \textbf{117}, 190501 (2016).
	\bibitem{421km} A. Boaron \textit{et al.} \textit{Phys. Rev. Lett.} \textbf{121}, 190502 (2018).
	\bibitem{free-space-QKD1} S.-K. Liao \textit{et al.} Nature \textbf{549}, 43 (2017).
	\bibitem{free-space-QKD2} H. Takenaka \textit{et al.} Nat. Photon. \textbf{11}, 502 (2017).
	\bibitem{Lucamarini-TF} M. Lucamarini, Z. L. Yuan, J. F. Dynes, and A. J. Shields. \textit{Nature} \textbf{557}, 400 (2018).
	\bibitem{Tamaki-security-TF} K. Tamaki, H.-K. Lo, W. Wang, and M. Lucamarini. \textit{preprint arXiv}:1805.05511.
	\bibitem{Ma-security-TF} X. Ma, P. Zeng, and H. Zhou. \textit{Phys. Rev. X} \textbf{8}, 031043 (2018).
	\bibitem{Cui-security-TF} C. Cui \textit{et al}. \textit{Phys. Rev. Appl.} \textbf{11}, 034053 (2019).
	\bibitem{Lutkenhaus-security-TF} J. Lin and N. L\"utkenhaus, \textit{Phys. Rev. A} \textbf{98}, 042332 (2018).
	\bibitem{Curty-security-TF} M. Curty, K. Azuma, and H.-K. Lo. \textit{npj Quantum Information} \textbf{5}, 64 (2019).
	\bibitem{Wang-security-TF} X.-Y. Zhou, C.-H. Zhang, C.-M. Zhang, and Q. Wang. \textit{Phys. Rev. A} \textbf{99}, 062316 (2019).
	\bibitem{Grasselli-Curty-TF} F. Grasselli and M. Curty. \textit{New J. Phys.} \textbf{21}, 073001 (2019).
	\bibitem{Grasselli-Navarrete-TF} F. Grasselli, A. Navarrete, and M. Curty. \textit{preprint arXiv}:1907.05256.
	\bibitem{experiment-chinese} Y. Liu \textit{et al.} \textit{preprint arXiv}:1902.06268.
	\bibitem{experiment-Toshiba} M. Minder, M. Pittaluga, G. L. Roberts, M. Lucamarini, J. F. Dynes, Z. L. Yuan, and A. J. Shields. \textit{Nat. Photon.} \textbf{13}, 334-338 (2019).
	\bibitem{experiment-Toronto} X. Zhong, J. Hu, M. Curty, L. Qian and H.-K. Lo. \textit{preprint arXiv}:1902.10209.
	\bibitem{experiment-Wang} S. Wang \textit{et al.} \textit{Phys. Rev. X} \textbf{9}, 021046 (2019).
	\bibitem{Takeoka} M. Takeoka, S. Guha, and M. M. Wilde. \textit{Nat. Comm.} \textbf{5}, 5235 (2014).
	\bibitem{PLOB} S. Pirandola, R. Laurenza, C. Ottaviani, and L. Banchi. \textit{Nat. Comm.} \textbf{8}, 15043 (2017).
	\bibitem{Epping} M. Epping, H. Kampermann, C. Macchiavello, and D. Bruß. \textit{New J. Phys.} \textbf{19}, 093012 (2017).
	\bibitem{Ribeiro} J. Ribeiro, G. Murta, and S. Wehner. \textit{Phys. Rev. A} \textbf{97}, 022307 (2018).
	\bibitem{Grasselli} F. Grasselli, H. Kampermann, and D. Bruß. \textit{New J. Phys.} \textbf{20}, 113014 (2018).
	\bibitem{Jo} Y. Jo and W. Son. \textit{OSA Continuum} \textbf{2}, 814-826 (2019).
	\bibitem{Pirandola-CV-CKA} C. Ottaviani, C. Lupo, R. Laurenza, and S. Pirandola. \textit{preprint arXiv}:1709.06988.
	\bibitem{Horodecki} R. Augusiak and P. Horodecki. \textit{Phys. Rev. A} \textbf{80}, 042307 (2009).
	\bibitem{Azuma-broadcast-network} S. Bäuml and K. Azuma. \textit{Quantum Sci. Technol.} \textbf{2} 024004 (2017).
	\bibitem{Wstate} W. Dür, G. Vidal, and J. I. Cirac. \textit{Phys. Rev. A} \textbf{62}, 062314 (2000).
	\bibitem{first-multiport-bs} M. Żukowski, A. Zeilinger, and M. A. Horne. \textit{Phys. Rev. A} \textbf{55}, 2564 (1997).
	\bibitem{multiport-beamsplitter} Y. L. Lim and A. Beige. \textit{Phys. Rev. A} \textbf{71}, 062311 (2005).
	\bibitem{not-balanced-transmittivities} A. Peruzzo, A. Laing, A. Politi, T. Rudolph, and J. L. O'Brien \textit{Nat. Comm.} \textbf{2}, 224 (2011).
	\bibitem{experimental-tritter} N. Spagnolo \textit{et al.} \textit{Nat. Comm} \textbf{4}, 1606 (2013).
	\bibitem{universal-multiport-interferferometers} W. R. Clements \textit{et al.} \textit{Optica} \textbf{3}, 1460-1465 (2016).
	\bibitem{linear-optics-for-4port-interferometer} G. N. M. Tabia	\textit{Phys. Rev. A} \textbf{86}, 062107 (2012).
	\bibitem{NV-centers-experiment} H. Bernien \textit{et al.} \textit{Nature} \textbf{497} pages 86–90 (2013).
	\bibitem{Wehner-experiment} F. Rozpedek \textit{et al.} \textit{Phys. Rev. A} \textbf{99}, 052330 (2019).
	\bibitem{long-coherence} M. H. Abobeih, J. Cramer, M. A. Bakker, N. Kalb, M. Markham, D. J. Twitchen, and T. H. Taminiau. \textit{Nat. Comm.} \textbf{9}, 2552 (2018).
	\bibitem{ScaraniRenner} V. Scarani and R. Renner. \textit{Phys. Rev. Lett.} \textbf{100}, 200501 (2008).
	\bibitem{Sheridan} L. Sheridan, T. P. Le, and V. Scarani. \textit{New J. Phys.} \textbf{12}, 123019 (2010).
	\bibitem{Tomamichel} M. Tomamichel, C. Lim, N. Gisin, and R. Renner. \textit{Nat. Comm.} \textbf{3}, 634 (2012).
	\bibitem{Curty-finiteMDI} M. Curty, F. Xu, W. Cui, C. Lim, K. Tamaki, and  H.-K. Lo. \textit{Nat. Comm.} \textbf{5}, 3732 (2014).
	\bibitem{tail-ineq} H.-L. Yin and Z.-B. Chen. \textit{preprint arXiv}:1903.09093.
	\bibitem{finite-key-TF} F.-Y. Lu \textit{et al.} \textit{preprint arXiv}:1901.04264.
	\bibitem{uncert-rel} M. Tomamichel and R. Renner. \textit{Phys. Rev. Lett.} \textbf{106}, 110506 (2011).
	
\end{thebibliography}
\end{document}